\documentclass{emulateapj}
\usepackage{amsmath}

\providecommand{\e}[1]{\ensuremath{\times 10^{#1}}}

\newcommand{\beq}[1]{\begin{equation}\label{#1}}
\newcommand{\eeq}{\end{equation}}
\newcommand{\unten}[1]{$_{\rm{#1}}$}
\newcommand{\tothe}[1]{$^{\rm{#1}}$}
\newcommand{\posion}[1]{$_{\rm{#1}}^{\rm{+}}$}

\newcommand{\Nperp}{N_{\perp}}

\newcommand{\CMC}{Chiang:2007p3804}
\newcommand{\DS}{Draine:1987p3323}
\newcommand{\BG}{Bai:2009p3370}
\newcommand{\IN}{Ilgner:2006p3319}
\newcommand{\OD}{Oppenheimer:1974p1573}

\shorttitle{PAHs, MRI, and Planets}
\shortauthors{Perez-Becker \& Chiang}

\begin{document}

\title{SURFACE LAYER ACCRETION IN TRANSITIONAL AND CONVENTIONAL DISKS:\\ FROM POLYCYCLIC AROMATIC HYDROCARBONS TO PLANETS}

\author{Daniel Perez-Becker\altaffilmark{}}
\affil{Department of Physics, University of California, Berkeley, CA 94720, USA}

\and

\author{Eugene Chiang\altaffilmark{}}
\affil{Departments of Astronomy and Earth and Planetary Science, University of California, Berkeley, CA 94720, USA}
\email{Electronic address: perez-becker@berkeley.edu}

\begin{abstract}

  ``Transitional'' T Tauri disks have optically thin holes with radii
  $\gtrsim 10$ AU, yet accrete up to the median T Tauri rate.
  Multiple planets inside the hole can torque the gas to high radial
  speeds over large distances, reducing the local surface density
  while maintaining accretion. Thus multi-planet systems, together
  with reductions in disk opacity due to grain growth, can explain how
  holes can be simultaneously transparent and accreting.  There
  remains the problem of how outer disk gas diffuses into the
  hole. Here it has been proposed that the magnetorotational
  instability (MRI) erodes disk surface layers ionized by stellar
  X-rays. In contrast to previous work, we find that the extent to
  which surface layers are MRI-active is limited not by ohmic
  dissipation but by ambipolar diffusion, the latter measured by $Am$:
  the number of times a neutral hydrogen molecule collides with ions
  in a dynamical time. Simulations by Hawley \& Stone showed
  that $Am \sim 100$ is necessary for ions to drive MRI turbulence in
  neutral gas. We calculate that in X-ray-irradiated surface layers, $Am$ typically
  varies from $\sim$$10^{-3}$ to 1,
  depending on the abundance of charge-adsorbing polycyclic aromatic
  hydrocarbons, whose properties we infer from {\it Spitzer}
  observations. We conclude that ionization of H$_2$ by X-rays and
  cosmic rays can sustain, at most, only weak MRI turbulence in
  surface layers 1--10 g/cm$^2$ thick, and that accretion rates in
  such layers are too small compared to observed accretion rates for
  the majority of disks.
\end{abstract}

\keywords{accretion, accretion disks --- instabilities --- ISM:
  molecules --- MHD --- planetary
  systems: protoplanetary disks --- stars: pre-main sequence}

\section{INTRODUCTION}
\label{intro}

On the road from molecular clouds to planetary systems,
transitional disks are among the brightest signposts.
Encircling T Tauri and Herbig Ae/Be stars having ages of 1--10 Myr, these disks
have large inner holes nearly devoid of dust. Identified by 
spectral energy distributions (SEDs; e.g., \citealp{Strom:1993p5527,Calvet:2005p2200,Kim:2009p5659}) 
and imaged directly (e.g., \citealp{Ratzka:2007p5686,Hughes:2007p2254,Brown:2009p6039}),
transitional disk cavities have radii on the order of 3--100 AU. Transitional
disks are so named
\citep{Strom:1990p6044}
because they might represent an evolutionary link
between optically thick disks without holes 
(e.g., \citealt{Watson:2007p6080})
and debris disks containing only rings of optically thin dust 
(e.g., \citealt{Wyatt:2008p6088}).
They are of special interest not least because their central clearings may
harbor nascent planets, potentially detectable against relatively weak
backgrounds.

\subsection{The Need for Companions}
\label{planets}

The idea that transitional disk holes are swept clean by
companions, possibly of planetary mass, is natural. We adhere to this
interpretation, although not all our arguments as given below are
the ones usually discussed.

\subsubsection{Stellar-mass Companions}
Roughly half of all transitional disks---6 out of 13 in the sample of
\citet{Kim:2009p5659}---are already known to contain stellar-mass
companions.  A prototypical example is CoKu Tau/4: a transitional
system whose hole is practically empty, both of dust
\citep{DAlessio:2005p6092} and gas (G.~Blake, private
communication, 2007). Inside its hole of radius $\sim$10 AU resides a nearly
equal mass K-star binary having a projected separation of 8 AU
\citep{Ireland:2008p6094}.  Gravitational torques exerted by the
binary can easily counteract viscous torques in the disk
\citep{Goldreich:1980p6130}, staving off accretion onto either star
\citep{Artymowicz:1994p6137,Ochi:2005p6170}.  Indeed stellar accretion
rates for CoKu Tau/4 are unmeasurably small, $\lesssim 10^{-10}
M_{\odot} \,{\rm yr}^{-1}$ \citep{Najita:2007p6175}.

Not every hole, however, is as empty as that of CoKu Tau/4. In many
cases there is a sprinkling of dust: optical depths at 10 $\mu$m
wavelength for many sources range from 0.01--0.1 
\citep{Calvet:2002p5431,Calvet:2005p2200}.
Observations of rovibrational emission from
warm, optically thick CO imply that gas fills many disk holes
\citep{Salyk:2007p6783}, albeit with surface densities that may be
far below those of conventional disks (we will argue below that this
is in fact the case). 
Most germane to our work, the host stars of many
transitional systems actively accrete, at rates that on average are
somewhat lower than those of conventional disks
\citep{Najita:2007p6175}, but which in several instances approach
$10^{-8} M_{\odot} \,{\rm yr}^{-1}$, the median T Tauri rate. The
holes of these systems must contain accreting gas.

How can we
reconcile the fact that many holes contain gas, accreting at rates
approaching those of conventional disks, with the fact that the holes
contain only trace amounts of dust?
To explain the paradox of simultaneous accretion and hole
transparency, appeals are sometimes made to grain growth, or the
filtering of dust out of gas by hydrodynamic mechanisms
\citep{Paardekooper:2006p6188,Rice:2006p6235}
or radiation pressure \citep{Chiang:2007p3804}.
These proposals, which invoke changes in disk opacity, may be part
of the solution. But they cannot alone explain the observations.
\citet{Ward:2009p6313} has criticized the hydrodynamic filter. 
The force of radiation pressure depends on uncertain optical constants and
grain porosities, and is likely to expel
only grains having a narrow range of sizes \citep{Burns:1979p7789}. 
Even if grains have the right properties to be blown out by
radiation pressure in vacuum, the inward flow of accreting gas may be
strong enough to carry as much as half of the grains that leak from
the rim into the hole \citep{Chiang:2007p3804}.
Grain growth does not explain why transitional disk accretion rates
$\dot{M}$ tend to be several times smaller than for
conventional T Tauri systems \citep{Najita:2007p6175}. Finally, none of these
proposals predicts gapped (``pre-transitional'') disks, which are
optically thick at stellocentric distances $a \lesssim 0.15$ AU (e.g.,
LkCa 15; \citealt{Espaillat:2007p5249}).

An alternative explanation for why holes can be simultaneously 
transparent and still contain accreting gas involves 
the special way in which disk gas accretes in the presence of
companions, particularly those on eccentric orbits.  If the hole rim
leaks gas---and for some combinations
of disk viscosity, disk pressure, and binary parameters the rim can be
quite leaky \citep{Artymowicz:1996p6337,Ochi:2005p6170}---the
gas can suddenly plunge inward at rates approaching freefall
velocities. The catastrophic loss of angular momentum is enabled by
the non-axisymmetric and time-dependent potential of the eccentric
binary, which directs gas streamlines onto radial orbits that may
intersect and shock. \citet{Artymowicz:1996p6337}
explained the paradox of simultaneous accretion and transparency:

\begin{quote}
`` $\ldots$ [for some
circumbinary disk parameters] gravitational resonant torques are able
to open a fairly wide gap [hole], while {\it concurrently} the
accretion flow proceeds through that gap in the form of
time-dependent, {\it well-developed} or {\it efficient} gas stream(s)
carrying virtually all the unimpeded mass flux. The radial velocity of
the stream is of order [the Kepler velocity], i.e., $\sim$$Re$
[Reynolds number] times faster than in the disk. By mass conservation,
the axially averaged surface density must differ by a factor of $Re >
10^3$ between the gap and the disk edge region. $\ldots$ The
spectroscopic ramification of this is a deficit of the observed
radiation flux emitted at temperatures appropriate for the gap
location.'' (italics theirs)
\end{quote}

Thus reductions in the dust-to-gas ratio by grain growth or dust
filtration are not the only processes that can render accreting gas transparent
in transitional disk holes. Companions can accelerate disk gas to such high radial speeds
that, by mass continuity, the surface density in {\it both} gas and dust is
reduced by orders of magnitude. According to this explanation, the
reduction in total surface density is not necessarily due to
consumption of gas by companions, but is rather due to
gravitational forcing.

\subsubsection{Multiple Planetary Mass Companions}
Though a stellar-mass companion can exert torques strong enough to
maintain holes of large size, a single planet-mass
companion on a circular orbit cannot do the same job. For
observationally reasonable values of the disk viscosity, a single
companion having of order $\sim$1 Jupiter mass embedded within a disk
at least a few times more massive carves out only a narrow gap (e.g.,
\citealt{Lubow:2006p6368}; \citealt{Crida:2007p6727}).  Disk gas
funnels past the planet by traveling on horseshoe-like orbits
\citep{Lubow:1999p6369}.
A single planet may siphon off some of the gas
that flows past it, but the disk accretion rate inside the planet's orbit
is reduced from that outside by a factor of $\lesssim 10$ 
\citep{Lubow:2006p6368}.
This modest reduction in $\dot{M}$, combined with the
narrowness of the gap seen in simulations ($\Delta r / r \sim 0.1$), implies inner disks far too
extensive and optically thick to explain transitional systems.

How, then, do planets fit in?  In those cases where the central stars
of transitional disks lack stellar-mass companions, the same tasks of
maintaining hole rims and increasing accretion velocities $v$ (but
not accretion rates $\dot{M}$) can be performed, not by a
single planet, but by a system of multiple planets.  We imagine a
series of planets, with the outermost lying just interior to and
shepherding the hole rim. Gas that leaks from the rim is torqued from
planet to planet, all the way down to the central star, its optical depth
decreasing inversely as its radial speed. The more massive the planets and the
more eccentric their orbits, the fewer of them should be required.

Such a picture is supported by numerical simulations of Jupiter
and Saturn embedded within a viscous disk \citep{Masset:2001p6788,Morbidelli:2007p6819}. 
In these simulations the two planets were close enough
that their gaps overlapped. Gas outside Saturn's orbit executed half a
horseshoe turn relative to Saturn, and then another half-horseshoe
turn relative to Jupiter, thereby crossing from the outer disk
through the Jupiter-Saturn common gap into the inner disk. 
\citet{Morbidelli:2007p6819} found that the surface density in the gap
region was reduced by 1--2 orders of magnitude, at least near Jupiter.

As our paper was being reviewed, we became aware of 
planet-disk simulations by Zhu et al.~(2010, submitted) which included
as many as 4 Jupiter-mass planets and whose results supported those of
\citet{Morbidelli:2007p6819}. Depending on the assumed efficiency with which planets
consumed disk gas, a set of four planets was found to reduce surface densities
in their vicinity by up to 2 orders of magnitude, while disk accretion rates
were reduced by factors $\lesssim 10$ (see their run P4A10). However, such surface density suppressions 
are not by themselves large enough to explain the observed low optical depths
of disk holes. Zhu et al.~(2010) concluded that reductions in gas opacity
by some means of dust depletion (e.g., grain growth) are still required.
Companions can also accommodate gapped or ``pre-transitional'' disks
in which optically thin holes contain optically thick annuli. As
inferred from spatially unresolved spectra, these annuli are narrow
and abut their host stars, extending mere fractions of an AU in radius
(\citealt{Espaillat:2007p5249}; but see also
\citealt{Eisner:2009p7325} who showed using spatially resolved
observations that the gapped disk interpretation of SR 21 is
incorrect).
In regions far removed from secondary companions---in particular,
in those regions closest to the primary star where the potential is
practically that of a point mass---the infall speeds of accreting gas
must slow back down to the normal rate set by disk viscosity.  By
continuity, the surface density must rise back up, and optical
thickness is thus restored.

\subsection{Companions are Not Enough: The Case for the Magnetorotational Instability for the Origin of Disk Viscosity}
\label{mri}

Our case for companions presumes a source of disk
viscosity. While a stellar-mass companion or a system of multiple
planets can transport gas quickly, effectively generating
an enormous viscosity in their vicinity
(i.e., inside the hole), they cannot cause the outer
disk to diffuse in the first place. An inviscid outer disk will not leak. Another source of
viscosity has to act in the outer disk, causing it bleed inward and
supply the observed accretion rates $\dot{M}$. We now turn to
the main subject of this paper, the possibility that the
magnetorotational instability is the source of viscosity in the outer disk.

The magnetorotational instability (MRI) amplifies magnetic fields in
outwardly shearing disks and drives turbulence whose Maxwell stresses
transport angular momentum outward and mass inward (for a review, see
\citealt{Balbus:2009p5490}).  Gas must be sufficiently well ionized
for the MRI to operate. For the most part, T Tauri and Herbig Ae disks
are too cold at their midplanes for thermal ionization to play a role
there. The hope instead is that X-rays emitted by host stars can
provide the requisite ionization in irradiated disk surface layers
\citep{Glassgold:1997p2130}.  The basic picture was conceived by
\citet{Gammie:1996p3339}, who proposed that disk surface layers
ionized by some non-thermal means may accrete, leaving behind
magnetically ``dead'' midplane gas.  Like other workers (e.g.,
\citealt{Bai:2009p3370}, hereafter BG; and \citealt{Turner:2010p6693},
hereafter TCS), we focus in this study on ionization of H$_2$ by
X-rays. Ionization of trace species by ultraviolet (UV) radiation is also
potentially important---we discuss this topic briefly at the close of
our paper.

The exposed rim of a transitional disk constitutes a kind of surface
layer. X-rays may penetrate the rim wall, activate the MRI there, and
dislodge a certain radial column of gas every diffusion time
\citep[][hereafter CMC]{\CMC}. Within the MRI-active column, both the
magnetic Reynolds number
\begin{equation}\label{Re}
Re \equiv \frac{c_{\rm s}h}{D} \approx 1 \left( \frac{x_{\rm
      e}}{10^{-13}} \right) 
\left( \frac{T}{100\, {\rm K}} \right)^{1/2} 
\left( \frac{a}{{\rm AU}} \right)^{3/2} 
\end{equation}
and the ion-neutral collision rate (normalized to the orbital frequency)
\begin{equation}\label{Am}
Am \equiv \frac{x_{\rm i} n_{\rm H_2} \beta_{\rm in}}{\Omega} \approx 1 \left( \frac{x_{\rm i}}{10^{-8}} \right) \left( \frac{n_{\rm H_2}}{10^{10}\, {\rm cm}^{-3}} \right) \left( \frac{a}{\rm AU} \right)^{3/2}
\end{equation}
must be sufficiently large for magnetic fields to couple well to the
overwhelmingly neutral disk gas. Here $T$ is the gas temperature,
$c_{\rm s}$ is the gas sound speed, $h = c_{\rm s}/\Omega$ is the gas
scale height, $\Omega$ is the Kepler orbital frequency, $D = 234 \,
(T/{\rm K})^{1/2} \, x_{\rm e}^{-1}$ cm$^2$ s$^{-1}$ is the magnetic
diffusivity, $x_{\rm e(i)}$ is the fractional abundance of electrons
(ions) by number, $n_{\rm H_2}$ is the number density of hydrogen
molecules, $\beta_{\rm in} \approx 1.9 \times 10^{-9}$ cm$^3$ s$^{-1}$
is the collisional rate coefficient for ions to share their momentum
with neutrals \citep{Draine:1983p6430}, and $a$ is the disk radius.

Dimensionless number (\ref{Re}) governs how well magnetic fields
couple to plasma, while (\ref{Am}) assesses how well plasma couples to
neutral gas. Both these numbers must be large for good coupling
between magnetic fields and neutral gas.
Numerical simulations 
have suggested critical values $Re^\ast$ of $\sim$$10^2$--$10^4$
\citep{Fleming:2000p6431},\footnote{The magnetic Reynolds number as we
  define it is not as accurate a predictor of MRI turbulence as the
  Elsasser (a.k.a.~Lundquist) number, which is given by (\ref{Re})
  with $c_{\rm s}$ replaced by the vertical Alfv\'en speed $v_{{\rm
      A}z}$. Self-consistent resistive MHD simulations by
  \citet{Turner:2007p3503} 
found that MRI-active regions coincide with Elsasser
  numbers greater than unity (see also \citealt{Sano:2001p7966}; \citealt{Sano:2002p7863}).
Our criterion $Re \gtrsim 10^2$--$10^4$ offsets some of
  the inaccuracy because $v_{{\rm A}z} \lesssim 10^{-1}
  c_{\rm s}$ in simulations of MRI turbulence. In any case we will
  find that the limiting factor for active surface layers is not $Re$ but
  rather $Am$.}  depending on the initial field geometry, and
$Am^{\ast}$ of $\sim$$10^2$ \citep[][hereafter HS]{Hawley:1998p5481}.  Some studies (e.g., TCS)
assumed $Am^{\ast} \sim 1$ in their determination of the
thicknesses of MRI-active surface layers, but numerical simulations of
marginally coupled ion-neutral systems indicated $Am^\ast$ may be 2
orders of magnitude higher (HS). The value of $Am^\ast$ is
critical to our work.

For typical T Tauri parameters, CMC found active radial column
densities $N^{\ast} \sim 5 \times 10^{23} \, {\rm cm}^{-2}$ or
equivalently mass columns of $\Sigma^\ast \sim 2 \, {\rm g} \, {\rm
  cm}^{-2}$---essentially the stopping column for 3 keV X-rays. When
they combined their derived value for $N^\ast$ with an assumed value for
the dimensionless disk viscosity $\alpha \sim 10^{-2}$, the accretion
rates of many transitional systems were successfully
reproduced. According to this model, the maximum accretion rate $\dot{M}$
inside the hole is set by conditions at the rim wall, i.e.,
by how large a radial column $N^\ast$ the MRI can draw from the
rim. Stellar or planetary companions, known or suspected to be present
(Section \ref{planets}), regulate how quickly this leaked material spirals
onto the host star---these companions modulate the radial inflow speed
$v(a)$ and thus the surface density $\Sigma(a) = \dot{M}/(2\pi
va)$.
But the companions inside the hole do not initiate disk accretion. They may reduce $\dot{M}$ by exerting repulsive torques to keep material
in the rim wall from leaking in, or by accreting material that flows
past \citep[e.g.,][]{Lubow:2006p6368,Najita:2007p6175}. But they
do not generate a non-zero $\dot{M}$ in the first place.
That fundamental task is
left to the MRI operating at the rim---or whatever source of anomalous
viscosity must be present in the outer disk to make it bleed.

\subsection{The Threat Posed by Polycyclic Aromatic Hydrocarbons to
  the MRI}

One concern raised by CMC but left quantitatively unaddressed is the
degree to which ultra-small condensates---macromolecules whose sizes
are measured in angstroms---may thwart the MRI. In planetary
atmospheres, aerosols can strongly damp electrical conductivities
(e.g., \citealp{Schunk:2004p6432,Borucki:2008p2667}).\footnote{Some
  fire alarms work on this principle. A radioactive source inside the
  alarm drives ionization currents in the air which normally complete
  an electrical circuit. When smoke particles from a fire
  reduce the density of free ions and electrons in air, the circuit is
  broken and the alarm is triggered.}
Most studies of active layers neglect aerosols and fixate on roughly
micron-sized grains, despite the fact that in many particle size
distributions, the smallest particles collectively present the
greatest geometric surface area and therefore the greatest cross
section for electron adsorption and ion recombination.  Exceptions
include \citet{Sano:2000p2094},
who in one model considered a grain size distribution extending down
to 0.005 $\mu$m = 50 $\AA$, and BG, who considered grain sizes as
small as 0.01 $\mu$m = 100 $\AA$.  Both studies found that in
principle small grains can be deadly to the MRI.

Notwithstanding their possibly decisive role, small grains are
sometimes wishfully dismissed as being depleted in number by grain
growth, i.e., assimilated into larger grains.  Undeniably grains grow
(\citealp{Blum:2008p6434,Chiang:2009p6460}),
so much so that their collective mass may be concentrated in particles
millimeters in size. But the question relevant for ionization
chemistry is not where the mass is weighted in the size spectrum of
particles, but rather where the collective surface area for charge
neutralization is weighted.
Determining the grain size distribution in disks seems a problem
that cannot be forward modeled with confidence.
\citet{Sano:2000p2094}
and BG instead parameterized the population of small grains and
studied the effects of varying their numbers, leaving undecided the
question of whether their parameter choices were favored by
observation or theory.

Like \citet{Sano:2000p2094} and BG, this paper considers
the effects of small condensates on the MRI.
What is new about our contribution is that we consider
the smallest imaginable condensates that are still accessible
to observation: polycyclic aromatic hydrocarbons (PAHs).
These molecules, typically containing
several dozens of carbon atoms, are excited electronically by
ultraviolet radiation and fluoresce vibrationally at 
3.3, 6.2, 7.7, 8.6, 11.3, and 12.7 $\mu$m,
the signature bands of their constituent C-C and C-H bonds
(e.g., \citealp{Li:2001p2937,Pendleton:2002p6462}).
{\it Spitzer} satellite spectra and ground-based adaptive optics imaging
reveal PAHs to be fluorescing strongly in Herbig Ae/Be and
T Tauri disk surface layers directly
exposed to stellar ultraviolet radiation \citep{Geers:2006p3349,
  Geers:2007p3824, Goto:2008p2626}.  Thus PAHs help to constrain the
aerosol abundance where magnetically driven accretion is
thought to occur: in disk surface layers.

In this work we incorporate PAHs into a simple chemical network to
assess the proposal that X-ray driven MRI operates in disk surface layers, either
on the top and bottom faces of conventional hole-less disks, or at the
rims of transitional disks. We make as realistic an estimate as we can
of the PAH abundance based on observations, to gauge how deep
the X-ray-irradiated, MRI-active layer might actually be.

To summarize this introduction: companions---either stars or a system
of multiple planets, but not a single Jupiter-mass planet---can help 
clear the extensive holes of transitional disks. The
outermost companion serves to establish the location of the rim where
viscous torques in the disk and gravitational torques from the
companion seek balance. If gas leaks inward from the outer disk, it is
driven onto the host star so quickly by gravitational torques from
companions that its optical depth may be reduced by orders of magnitude.
Companions, together with reductions in disk opacity by grain growth, thus maintain the transparency of
the hole while still permitting stars to accrete gas. But
companions do not, in and of themselves, cause gas in the outer disk
to diffuse inward. That responsibility may be reserved for the
MRI---whose ability to operate despite the presence of
charge-neutralizing PAHs is the subject of this paper.

Our paper is organized as follows. The ingredients of our numerical
model for X-ray-driven ionization chemistry in disk surface layers are
laid out in Section \ref{sec_model}. There we gauge what PAH abundances in
disks may be.  Results---principally, how $Am$ and $Re$ vary with the
column density penetrated by X-rays, and the extent to which PAHs
reduce these numbers---are presented in Section \ref{sec_results}.  Analytic
interpretations of our numerical results, and direct comparison with
previous calculations (BG, TCS), are given there as well. We discuss
our main results for X-ray driven MRI in Section \ref{sec_discussion}, and
close by discussing the possibility of UV-driven MRI.

\section{MODEL FOR DISK IONIZATION} 
\label{sec_model}

In this paper we are interested in the degree to which stellar X-rays
and Galactic cosmic-rays can ionize H$_2$ gas in T Tauri disks.  In
this respect our study is similar to many others, and we make direct
comparisons of our work to BG and TCS in Section \ref{compare}.
For simplicity our model neglects ionization of trace species like C and S
by stellar UV radiation. Omitting UV-driven chemistry
renders our model inconsistent because our model also includes PAHs,
whose abundances we constrain in Section \ref{sec_PAHs} by using
observed PAH emission lines excited by stellar UV radiation.
We will discuss the critical issue of UV ionization in Section \ref{future}.

\begin{figure} [ht]
\epsscale{1.0}
\plotone{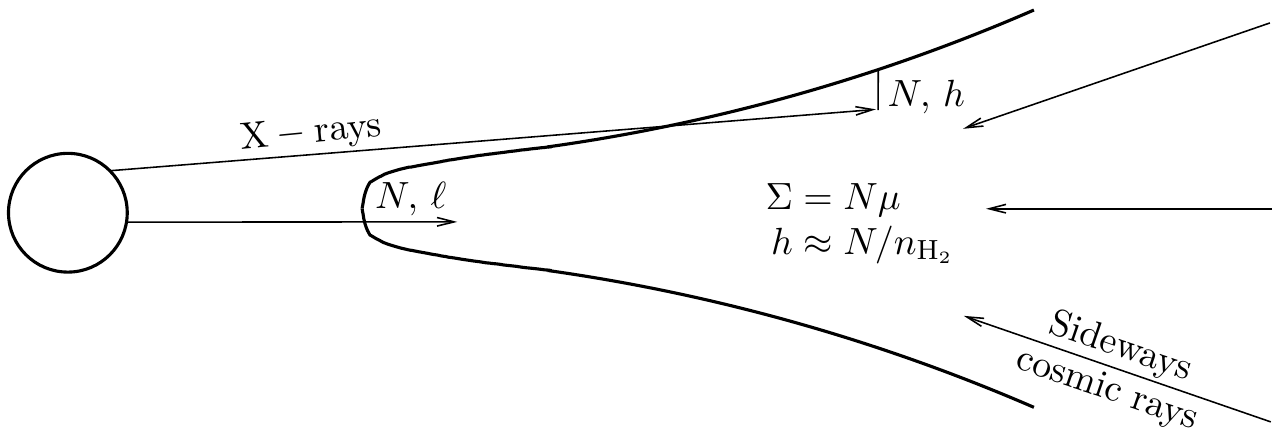}
\caption{X-ray ionized surface layers, located either on
  the top and bottom faces of a flared disk, or at its inner rim. Our
  calculations apply to both situations, although they are more
  accurate for the former. We assume in this work that the lengthscale
  $\ell$ over which gas is distributed radially at the hole rim is
  equal to $h$, the vertical gas scale height. Other sources of
  ionization are interstellar cosmic-rays and ultraviolet radiation
  from the star. At large stellocentric distances ($a \gtrsim 30$ AU), cosmic
  rays may penetrate the disk from the side. For most of our paper, we
  neglect ionization of trace species by far ultraviolet radiation,
  but in Section \ref{future} we briefly discuss this important topic.
} 
\label{disk_diagram}
\end{figure}

\subsection{X-ray and Cosmic-ray Ionization Rates and Gas Densities} 
\label{sec_ionize}

\textit{Chandra} spectra of pre-main-sequence stars in the Orion
Nebula can be fitted by a pair of thermal plasmas with characteristic
temperatures $kT_{\rm{X}} \sim 1$ and 3 keV, where $k$ is Boltzmann's
constant, and comparable luminosities $L_{\rm{X}} \sim
10^{28}$--$10^{31}$ erg s\tothe{-1}
\citep{Wolk:2005p4655,Preibisch:2005p6831}.  The softer component is
believed to be emitted by shock-heated accreting gas
\citep{Stelzer:2004p4742}, and the harder component by a strongly
magnetized and active stellar corona \citep{Wolk:2005p4655}.  X-ray
luminosities tend to increase with increasing stellar mass and
decreasing accretion rate (see Figure 1 of \citealt{Telleschi:2007p7899}, 
and Figure 17 of \citealt{Preibisch:2005p6831}). These correlations are
relevant for transitional disks because some transitional disks are
hosted by higher mass Herbig Ae stars, and accretion rates for
transitional disks tend to be lower than for conventional disks
\citep{Najita:2007p6175}. For our standard model we will adopt
$L_{\rm{X}} = 10^{29}$ erg s$^{-1}$, but we will also experiment with
$L_{\rm{X}} = 10^{31}$ erg s$^{-1}$. We fix the temperature of the
X-ray emitting plasma at $kT_{\rm{X}} = 3$ keV, an assumption that
ignores how X-ray spectra harden with increasing $L_{\rm X}$
\citep{Preibisch:2005p6831}. Although none of our numerical models
explicitly considers $kT_{\rm X} > 3$ keV, we will discuss
quantitatively in Section \ref{sec_lx} how our results scale with the higher
ionization rates afforded by a harder ($kT_{\rm X} = 8$ keV) X-ray
spectrum; we will see there that the effects are not large.\footnote{ A
  minority of sources surveyed by {\it Chandra} exhibited superhot
  X-ray flares with peak $L_{\rm X} \sim 10^{32}$ erg s$^{-1}$ and
  $kT_{\rm X} \sim 15$ keV (\citealt{Getman:2008p7904}; \citealt{Getman:2008p7905}).
The degree to which superhot flares enhance ionization rates depends on the
  uncertain flare duty cycle. Because only $\sim$10\% of the {\it
    Chandra} sources flared once or twice over a 15-day observing
  period, and because each flare lasted less than $\sim$1 day, the
  extra ionization from superhot flares may be modest. }  See Table
\ref{table_parameters} for a list of all model parameters.

We derive X-ray ionization rates $\zeta_{\rm X}$ as a function of penetration column
$N$ from \citet[][hereafter IG]{Igea:1999p2931}, who constructed a Monte
Carlo radiative transfer model that accounts for Compton scattering
and photoionization.  Compton scattering enables X-ray photons to
penetrate to deeper columns than would otherwise be possible.
For our standard model we use IG's Figure 3 for a thermal plasma of 
$L_{\rm{X}} =10^{29}$ erg s\tothe{-1} and $kT_{\rm{X}} = 3$
keV. Their ionization rates
were computed for stellocentric distances of 5 and 10 AU; we scale
these rates to the stellocentric distances of our model, $a = 3$ and
30 AU, using the geometric dilution factor $a^{-2}$. We test
the accuracy of this approach by scaling their results internally
using this dilution factor, finding (by necessity) excellent agreement
at low columns where material is optically thin to X-rays, and
agreement better than a factor of two at the highest columns
calculated.  For the case $L_{\rm{X}} = 10^{31}$ erg s\tothe{-1}, we increase
all ionization rates from our standard values by a factor of 100.

Interstellar cosmic-rays can also ionize disk gas, but are attenuated by
magnetized stellar winds blowing across disk surface layers.
Even the contemporary solar wind, characterized by a mass loss rate of $\sim$$10^{-14}
M_{\odot}$ yr$^{-1}$, modulates the cosmic-ray flux at Earth by as
much as $\sim$10\% with solar cycle \citep{Marsh:2003p6841}.  T
Tauri winds, having mass loss rates up to 5 orders of magnitude higher
than that of the solar wind today, seem likely to shield disk surfaces
from cosmic-rays directed normal to the disk plane
(cf.~\citealt{Turner:2009p6844}). Nevertheless, cosmic-rays may reach
disk gas from the ``side,'' striking the disk edge-on from the outside.
At $a = 3$ AU we estimate that these ``sideways cosmic-rays'' are too strongly
attenuated by intervening disk gas to be significant. The same
is not true on the outskirts of the disk at $a = 30$ AU, where column
densities measured radially outward may be smaller than the
cosmic-ray stopping column of 96 g cm$^{-2}$ \citep{Umebayashi:1981p7866}. 
Thus we consider another model at $a = 30$ AU where in addition to our
standard X-ray source we include sideways cosmic-rays with a constant,
column-independent ionization rate of $\zeta_{\rm{CR}} \sim (1/4)
\times 10^{-17}$ s$^{-1}$ (\citealt{Caselli:1998p3151}, hereafter C98). The factor of
$1/4$ is approximately the fraction of the celestial sphere (centered
on disk gas at 30 AU) that is not shielded by stellar winds.
The total ionization rate $\zeta = \zeta_{\rm X} + \zeta_{\rm CR}$.

In all our simulations we neglect ionization by energetic protons
emitted by the stars.
As discussed by \citet{Turner:2009p6844}, estimates of the
stellar proton flux rely on extrapolated scaling relations, and the
ability of particles to reach the disk surface in the face of strong
stellar magnetic fields is uncertain. Moreover, protons are emitted in
flares which may occur too infrequently to sustain disk ionization.
In one of their models, \citet{Turner:2009p6844} used a time-steady
stellar particle luminosity whose ionization rate exceeded, by a factor of 40 at a mass column of
$\Sigma = 8$ g cm$^{-2}$,
that of an X-ray source having $L_{\rm{X}}
= 2 \times 10^{30}$ erg s$^{-1}$ and $kT_{\rm X} = 5$ keV.
This model probably yields a hard upper limit on the stellar proton ionization
rate, derived under a set of generous assumptions.  Our $L_{\rm
  X} = 10^{31}$ erg s$^{-1}$ case produces ionization rates $\zeta$
that approach those of the aforementioned model 
to within an order of magnitude. In any case we will see in Section \ref{sec_lx}
how our results can be scaled to any $\zeta$.

Figure \ref{disk_diagram} depicts schematically how X-rays irradiate
disk surface layers, usually pictured in the vertical direction as
ensheathing the disk on its top and bottom faces. But in a
transitional disk, a surface layer may also be present in the radial
direction, at the rim of the central hole.  We consider each of these
environments in turn, estimating local number densities $n_{\rm H_2}
\, [\mathrm{H}_2 \, \mathrm{cm}^{-3}]$ from the column density $N \,
[\mathrm{H}_2 \, \mathrm{cm}^{-2}]$ penetrated by X-rays.\footnote{In
  this paper, ionization rates $\zeta$, column densities $N$, and
  fractional densities $x$ are referred to hydrogen molecules, not
  hydrogen nuclei.\label{factor2}}

\begin{deluxetable*}{llll}
\tablecaption{Model parameters.\label{table_parameters}}
\tablehead{\colhead{Parameter} & \colhead{Variable} & \colhead{Value} & \colhead{Reference}} \\
\startdata
Disk radius & $a$&3, 30 AU& \dots \\
X-ray source luminosity$^a$ & $L_{\rm{X}}$ &$10^{29} \,(10^{31})$ erg s$^{-1}$&Section  \ref{sec_ionize}  \\
X-ray source temperature & $kT_{\rm{X}}$&3 keV&Section \ref{sec_ionize} \\
Cosmic-ray ionization rate& $\zeta_{\rm{CR}}$ &0, $(0.25 \times 10^{-17})$
s$^{-1}$&\citet{Caselli:1998p3151}\\
Initial CO abundance$^{b}$ & $x_{\mathrm{CO}} $&$10^{-4}$& \citet{Aikawa:1996p3361}\\
Total metal abundance$^{a,b}$ & $x_{\rm M}$& $10^{-8} \,(0,\, 10^{-6})$&Section \ref{sec_metals} \\
Total grain abundance$^{b}$& $x_{\rm grain} $&$6 \times 10^{-15} \epsilon_{\rm
  grain}$&Section \ref{sec_grains}\\
Grain settling (depletion) factor&$\epsilon_{\rm grain}$&$10^{-3} \leq \epsilon_{\rm
  grain} \leq 10^{-1}$&Section \ref{sec_grains} \\
Total PAH abundance$^{b}$&$x_{\mathrm{PAH}}$ &$10^{-6} \epsilon_{\rm PAH}$& Section \ref{sec_PAHs}\\
PAH depletion factor&$\epsilon_{\rm PAH}$&$10^{-5} \leq \epsilon_{\rm
  PAH} \leq 10^{-2}$&Section \ref{sec_PAHs} \\
Central stellar mass& $M_{\ast}$&1 $M_{\sun}$& \dots\\  
Gas temperature&$T$&80, 30 K&Section \ref{sec_temp} \\
\enddata
\tablenotetext{a}{Values in parentheses correspond to test cases different from our standard model.}
\tablenotetext{b}{All abundances are relative to H$_2$ by number.}
\end{deluxetable*}

\subsubsection{Surface Layers I: Top and Bottom Faces of a Conventional Flared Disk}
\label{sec_surf1}

When considering the surface layers of a conventional non-transitional
disk, we describe our results
as a function of the vertical column density $N$ of hydrogen molecules,
measured perpendicular to and toward the disk midplane.
Thus our $N$ coincides with $\Nperp$ of IG, save for a factor
of 2 because IG count hydrogen nuclei whereas we count
hydrogen molecules. An equivalent measure of vertical column density
$N$
is the mass surface density $\Sigma \equiv N \mu$, where $\mu \approx 4
\times 10^{-24}$ g is the mean molecular weight of gas.

To good approximation, the local number density
\beq{nNa} n_{\rm H_2} \approx N/h \eeq
where the vertical scale height $h = c_{\rm s}/\Omega = (kT/\mu)^{1/2}/\Omega$.
For %
gas temperature $T \approx 80 (30)$ K at $a = 3 (30)$ AU
(see Section \ref{sec_temp} for how we derive these temperatures), we find $h = 0.09\, (1.8)$ AU.

Equation (\ref{nNa}) underpins all our calculations of chemical equilibrium.
For a typical $N \sim 10^{23}$ H$_2$ cm$^{-2}$, we have $n \sim 7 \times 10^{10} \, (4 \times 10^{9})$ H$_2$ cm$^{-3}$ at $a = 3 \,(30)$ AU.

\subsubsection{Surface Layers II: Gap Rim of Transitional Disk}
\label{sec_surf2}
We assume that the gap rim is not shadowed from the star by gas interior to
the rim. We cannot prove that the rim is not shadowed, but disk models
based on the infrared SED suggest it is not
\citep[e.g.,][]{Calvet:2005p2200}.
Possibly gas at the rim wall
``puffs up'' because it is heated by X-rays and can maintain
a larger vertical height than gas inside the hole (e.g., \citealt{Dullemond:2001p7074}).

For the case of the rim of a transitional disk, we 
reinterpret $N$ (equivalently $\Sigma$) as the radial column of hydrogen
molecules traversed by X-rays (Figure 1). To 
estimate the local number density $n_{\rm H_2}$,
we need to know the radial lengthscale $\ell$ over which
material at the rim wall is distributed. Plausibly $h \lesssim \ell \lesssim a$.
\citet{Chiang:2007p3804} take $\ell \sim a$,
but models based on SEDs and images suggest the rim is much sharper.
Here we assume that $\ell \sim h$ so that Equation (\ref{nNa})
applies equally well to transitional disks as to
conventional disks---keeping in mind
that $N$ should be measured radially for the former
and vertically for the latter.

To calculate ionization rates in the rim, we still use the results of
IG, reinterpreting their $\Nperp$ in their Figure 3 as our radial
column $N$.  Clearly the scattering geometry differs between the case
of a transitional disk rim and the case of the top and bottom faces of
IG's conventional disk.  Where material is optically thin to stellar
X-rays, the two cases match in ionization rate (per molecule), but
where it is optically thick, we underestimate the ionization rate 
in transitional disk rims by
using IG because more X-rays escape by scattering vertically out of conventional disk
surface layers than from the rim. Another reason we underestimate the
ionization rate at high column density is because the total X-ray flux
per unit surface area of the disk is lower for conventional
surface layers---which are illuminated at grazing incidence---than for
the rim, which is illuminated at normal incidence. Nevertheless we
estimate that these errors are of the order of unity, insofar as the
columns that might possibly be MRI-active are not too optically thick
to X-rays (Section \ref{sec_AmRe}), because the Thomson scattering
phase function is fairly isotropic, and the cross-section for
scattering is only comparable to that for photoionization at the relevant
photon energies. In any case we will explore the
effects of higher ionization rates by running a model with higher
$L_{\rm{X}} = 10^{31}$ erg s$^{-1}$ (Section \ref{sec_lx}).

\subsection{Gas Temperature}
\label{sec_temp}

The surface layers of protoplanetary disk atmospheres vary widely in
temperature, from $\sim$5000 K at the lowest columns where stellar
X-rays heat the gas, down to $\lesssim$ 100 K at the highest columns where
dust reprocesses optical starlight. We draw our temperatures from the 
thermal model of \citet{Glassgold:2004p2132}, which in turn is based
on the dust temperature model of \citet{DAlessio:1999p6620}.  At $a =
1$ AU at column densities of interest ($N \gtrsim 10^{22}$ cm$^{-2}$),
thermal balance is controlled primarily by reprocessing of
starlight by dust, and gas and dust temperatures are nearly equal at $\sim$130
K (\citealt{Glassgold:2004p2132}, their Figure 2). We adjust this
result for the disk radii of our standard model using the dust
temperature scaling law for the midplane of a passive flared disk, $T
\propto a^{-3/7}$ \citep[e.g.,][]{Chiang:1997p403}. Thus at $a = 3$
AU we have $T = 80$ K, and at $a = 30$ AU we have $T = 30$ K. Note
that these temperatures are lower---and arguably more realistic---than
those assumed by BG and TCS, who invoked
temperatures of the traditional Hayashi nebula without justification.

\subsection{Chemical Network}
\label{sec_chemnet}  
Following \citet[][hereafter IN]{Ilgner:2006p3319}, CMC, and BG, we apply a simple network of chemical
reactions based on that designed for molecular clouds by
\citet[][hereafter OD]{\OD}. \citet{\IN} and BG compared the results of
OD-based schemes to those of more complex networks extracted from the
UMIST (University of Manchester Institute of Science and Technology;
\citealt{Woodall:2007p4675}, hereafter W07; \citealt{Vasyunin:2008p6845}) database. Fractional
electron abundances derived by IN using the
simple network were greater than those derived using the complex
network, whereas BG, who used a more recent version of the UMIST
database, found that the sign of the difference varied from case to
case.  The magnitude of the difference ranged up
to a factor of 10, but was often $\lesssim$ 3.
Using the simple network seems the most practical approach, if we are
content with order-of-magnitude answers. 
In Section \ref{compare}, we test the results
of our code against those of BG and TCS.

\begin{figure*} [h!]
\epsscale{1}
\plotone{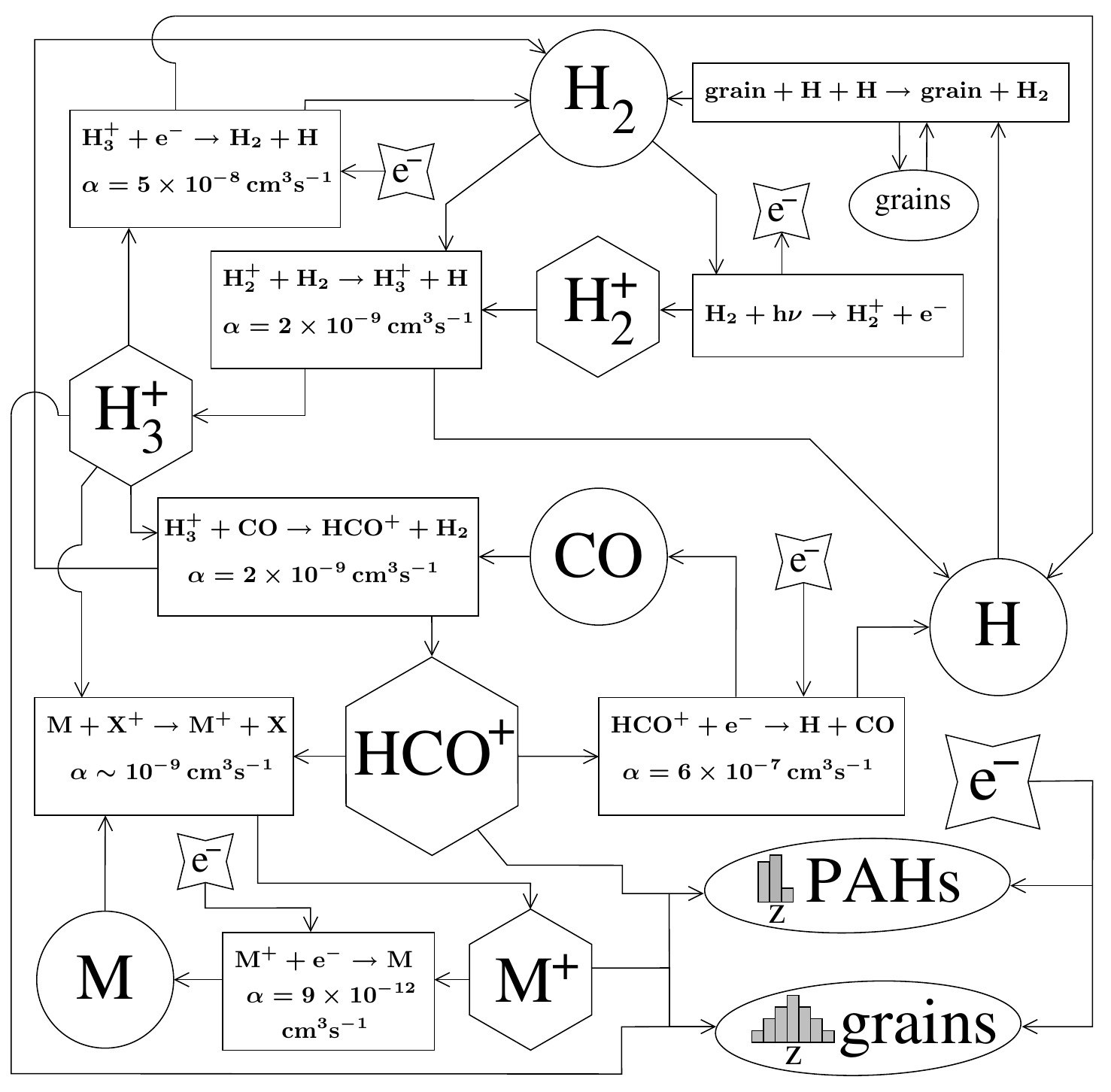}
\caption{Our chemical reaction network, derived from \citet{\OD}. Rate
  coefficients as shown in this Figure are evaluated at $T=80$ K. See
  Table \ref{table_eq} for a comprehensive list of all modeled
  reactions and precise rate coefficients.}\label{fig_reaction_network}
\end{figure*}

All reactions in our OD-based network are listed in Table \ref{table_eq}
and shown schematically in Figure \ref{fig_reaction_network}. 
Rate coefficients and their temperature dependences are taken
from the UMIST database. The chain of
events basically proceeds as follows. X-rays ionize H\unten{2} to
H\posion{2}, which rapidly reacts with H\unten{2} to produce
H\posion{3}. The H\posion{3} ion combines with CO to form
HCO\tothe{+}.  Most HCO\tothe{+} ions dissociatively recombine with
free electrons, but some transfer their charge to gas-phase metal
atoms such as Mg. Free metals tend to be abundant positive charge
carriers, as they recombine with free electrons only by a slow
radiative channel. Charged particles in the network (e\tothe{-},
H\posion{3}, HCO\tothe{+}, metal\tothe{+}) can neutralize by
collisionally transferring their charge to PAHs and grains.
The collisional charging process is described in Section \ref{sec_cc}.

The reaction loop is closed by the formation of H$_2$ on grain
surfaces. To compute the rate of this reaction, we take neutral H atoms
to collide with grains using the geometrical cross section for grains,
and adopt from BG the uniform probability $\eta = 10^{-3}$ for
a pair of adsorbed hydrogen atoms to form a hydrogen molecule (see
their Equation 27). The precise rate of this reaction is not important
for us, as it only sets the equilibrium abundance of H, which is
irrelevant for the ionization fraction, as long as $n_{\rm H} \ll
n_{\rm H_2}$.

In their original study OD included ionization of He and reactions
involving atomic and molecular oxygen. We neglect these for simplicity.
Most reactions involving oxygen initiate with the formation
of the hydroxyl ion (H\posion{3} + O $\rightarrow$ OH$^+$),
which proceeds at a rate only comparable to the formation
of HCO\tothe{+}, which we do account for. Thus
our neglect of oxygen within the OD framework is not expected to alter our
results for the fractional ionization by more than a factor of 2.
In any case, in Section \ref{compare}
we will compare our results with those of more complex
networks considered by BG and TCS.

\begin{deluxetable*}{lrcllll}
\tabletypesize{\tiny}
\tablewidth{0pt}
\tablecaption{Chemical Reactions Including Collisional Charging of PAHs and Grains.\label{table_eq}}
\tablewidth{0pt}
\tablehead{
\colhead{Number} &  & \colhead{Reaction}  &
  & \colhead{Rate Coefficient$^{\rm a}$} &\colhead{Value} & \colhead{Reference} 
}
\startdata
1$^{\rm b}$ & H\unten{2} + $h \nu$  &$\rightarrow$ &  H\posion{2}  +  e\tothe{-}& $\zeta_{\rm X}$ & Taken from radiative transfer model & IG\\
2 & H\unten{2} + Cosmic-ray  &$\rightarrow$ &  H\posion{2}  +  e\tothe{-}& $\zeta_{\rm CR}$ & $(1/4) \times 10^{-17}$ s$^{-1}$ & C98\\
3 & H\posion{2}  +  H\unten{2}  &$\rightarrow$ &  H\posion{3}  + H&$\alpha_{\mathrm{H_2^+,H_2}}$&2.1\e{-9} & W07\\
4 & H\posion{3} + CO &$\rightarrow$ & HCO\tothe{+} + H\unten{2}&$\alpha_{\mathrm{H_3^+,CO}}$&1.7\e{-9} & W07\\
5$^{\rm c}$ &M + H\posion{3} &$\rightarrow$ & M\tothe{+} +H\unten{2}
+ H&$\alpha_{\mathrm{M,X^+}}$&1.0\e{-9}& W07 \\
6&M + HCO\tothe{+}  &$\rightarrow$ & M\tothe{+}  + H +
CO&$\alpha_{\mathrm{M,X^+}}$&2.9\e{-9}& W07 \\
7 & H\posion{3} + e\tothe{-}  &$\rightarrow$ &  H\unten{2} +
H&$\alpha_{\mathrm{H_3^+,e}}$&2.3\e{-8} $(T/300\rm{K})^{-0.52}$ & W07  \\
8 &HCO\tothe{+} + e\tothe{-} &$\rightarrow$ &  H +
CO&$\alpha_{\mathrm{HCO^+,e}}$&2.4\e{-7} $(T/300\rm{K})^{-0.69}$& W07\\
9 &M\tothe{+} + e\tothe{-} & $\rightarrow$ & M +  
$h\nu$&$\alpha_{\mathrm{M^+,e}}$&2.8\e{-12} $(T/300\rm{K})^{-0.86}$& W07\\
10 &PAH(Z) + e\tothe{-} & $\rightarrow$ & PAH(Z$-1$) &$\alpha_{\mathrm{PAH,e}}$&Section  \ref{sec_cc} &DS\\ 
11$^{\rm d}$ &PAH(Z) + X\tothe{+} & $\rightarrow$ & PAH(Z+1) &$\alpha_{\mathrm{PAH,X^+}}$&Section  \ref{sec_cc} &DS\\ 
12 &grain(Z) + e\tothe{-} &$\rightarrow$ & grain(Z$-1$) &$\alpha_{\mathrm{grain,e}}$&Section  \ref{sec_cc} &DS\\ 
13 &grain(Z) + X\tothe{+} & $\rightarrow$ & grain(Z+1)  &$\alpha_{\mathrm{grain,X^+}}$&Section  \ref{sec_cc} &DS\\ 
14 &PAH(Z=$-1$) + PAH(Z=1) & $\rightarrow$ & $2 \times$ PAH(Z=0) &$\alpha_{\mathrm{PAH,PAH}}$& Section  \ref{sec_cc} &DS\\ 
15 &H + H + grain & $\rightarrow$ & H\unten{2} + grain &$\alpha_{\mathrm{physisorption}}$&Section  \ref{sec_chemnet} &BG\\ 
\enddata
\tablenotetext{a}{$\zeta$ has units of s$^{-1}$ and $\alpha$ has units
  of cm$^{3}$s$^{-1}$.}
\tablenotetext{b}{$h\nu$ denotes a photon.}
\tablenotetext{c}{M represents a gas-phase atomic metal, e.g., Mg.}
\tablenotetext{d}{X\tothe{+} can be either H\posion{3},
  HCO\tothe{+}, or M\tothe{+}.}
\end{deluxetable*}

\subsection{Properties and Abundances of Trace Species}\label{sec_trace}
The trace ingredients of our model include gas-phase metals
(Section \ref{sec_metals}), a monodispersion of micron-sized grains
(Section \ref{sec_grains}), and PAHs (Section \ref{sec_PAHs}).

\subsubsection{Gas-phase Metals (Magnesium)}
\label{sec_metals}
For gas-phase metals which serve importantly as electron donors (OD; \citealt{Fromang:2002p7072}),
we are guided by Mg, whose solar abundance is $3.5 \times 10^{-5}$ atoms per hydrogen
nucleus \citep{Lodders:2003p5309}. The fraction of Mg that is in the
gas phase---neither incorporated into grain interiors nor adsorbed
onto grain surfaces---might be at most 3--30\% by number, its value in
the diffuse interstellar medium \citep{Jenkins:2009p6986}.  In the dense
environments of protoplanetary disks, the gas-phase fraction should be
much smaller because magnesium is used toward building grains.

Nominally, our model temperatures of 30--80 K are so low that almost
all of the Mg not incorporated into grain interiors should be adsorbed
onto grain surfaces, leaving behind only a tiny fraction in the gas
phase (\citealt{Turner:2007p3503}, their Section 2.2; see also
Equation 26 of BG).  Just how tiny is uncertain, given how sensitive
the adsorption fraction is to gas temperature, and how steep
temperature gradients can be in disk surface layers
\citep{Glassgold:2004p2132}.  Turbulent mixing of hot, high altitude,
normally metal-rich layers with cold, low altitude, normally metal-poor
layers can also complicate matters (\citealt{Turner:2007p3503}; TCS).

We adopt a standard metal abundance of $x_{\rm{M}} = 10^{-8}$ metal
atoms per H$_2$, which corresponds to a gas-phase fraction of
$\sim$$10^{-4}$ by number relative to solar.  Our choice is similar to those of
IN and BG.
We also experiment with a metal-free case in which all metals have
been adsorbed onto grain surfaces ($x_{\rm M} = 0$), and a metal-rich
case for which $x_{\rm M} = 10^{-6}$. Although the
metal-rich case is not especially realistic and is not justified by
our model parameters---in particular our low gas temperatures---we
consider it anyway because we would like to understand the effects of
metals in principle, and to connect with other studies that consider
similarly large metal abundances (CMC; \citealt{Turner:2007p3503};
TCS).

\subsubsection{Grains}
\label{sec_grains}

The number of grains per H$_2$ molecule is 
\beq{grain_abundance} x_{\rm grain} =
\frac{\mu}{\frac{4}{3}\pi s^3 \rho_{s}} \frac{\rho_{\rm
    dust}}{\rho_{\rm gas}} \,, \eeq
\noindent where $\mu \approx 4 \times 10^{-24}$ g is the mean
molecular weight of gas
and $\rho_{\rm dust}/\rho_{\rm gas}$ is the
dust-to-gas mass ratio.
For simplicity we consider grains of a single radius $s = 1 \,\mu$m
and internal density $\rho_s = 2 \,{\rm g} \,{\rm cm}^{-3}$.  There is
ample evidence that micron-sized grains abound in surface layers,
both from mid-infrared spectra of silicate emission lines
\citep[e.g.,][]{Natta:2007p6483}
and from scattered light images at similar
wavelengths \citep[e.g.,][]{McCabe:2003p6487}.

The dust-to-gas ratio in surface layers may differ considerably
from its value in the well-mixed diffuse interstellar medium (ISM):
\beq{epsilon} \frac{\rho_{\rm dust}}{\rho_{\rm gas}} \equiv
\epsilon_{\rm grain} \left. \frac{\rho_{\rm dust}}{\rho_{\rm gas}}
\right|_{\rm ISM}  \,,
\eeq
where for the ISM of solar abundance $\rho_{\rm dust}/\rho_{\rm gas}|_{\rm ISM} = 0.015$ \citep{Lodders:2003p5309}.
Based on model fits to observed far-infrared SEDs
\citep[][]{Chiang:2001p5086,DAlessio:2006p5245,Dullemond:2004p7008},
there is consensus that surface layer grains directly
illuminated by optical light from their host stars
have settled toward the midplane into regions of denser gas.
Thus $\epsilon_{\rm grain} < 1$, but actual values
are not known with certainty, because small changes in the SED
resulting from small changes in the height of the dust photosphere
imply large changes in gas density in a near-Gaussian atmosphere.
For example, Figure 15 of \citet{DAlessio:2006p5245}
shows that changing the far-infrared SED by less than a factor of 2
changes $\epsilon_{\rm grain}$ by a factor of 10.

Table \ref{table_grains} lists fitted values of $\epsilon_{\rm grain}$
for some transitional disks, drawn from the literature. At best they are
accurate to order of magnitude.  For our calculations
we consider $10^{-3} \leq \epsilon_{\rm grain} \leq 10^{-1}$ (Table
\ref{table_parameters}).

\begin{table}[htb!]
\caption{Dust Settling Parameter $\epsilon_{\rm grain}$ for
  Some Transitional Disks}\label{table_grains}
\begin{tabular}{lrl}
\tableline\tableline
Source & $\epsilon_{\rm grain}$ & Reference \\
\tableline
LkCa 15&$10^{-3}$&\citet{Espaillat:2007p5249, Chiang:2001p5086}\\
UX Tau A&$10^{-2}$&\citet{Espaillat:2007p5249} \\
CS Cha&$10^{-2}$&\citet{Espaillat:2007p5236}\\
GM Aur&$10^{-1}$&\citet{Calvet:2005p2200}\\
DM Tau&$10^{-1}$&\citet{Calvet:2005p2200}\\

\end{tabular}
\end{table}

\subsubsection{PAHs}
\label{sec_PAHs}

For simplicity we model PAHs as spheres, each having a radius $s = 6
\AA$ and internal density $\rho_s = 2$ g cm$^{-3}$.  Although in
reality carbon atoms in PAHs are arranged in sheets and not spheres
\citep[e.g.,][]{Allamandola:1999p3344}, the difference in cross
section arising from geometry is only on the order of unity.  Each of
our model PAHs has about as much mass as a real PAH containing $N_{\rm
  C} = 100$ carbon atoms.  A PAH of this size is estimated to be just large
enough to survive photo-destruction around Herbig Ae stars
\citep{Visser:2007p3815}.

The central wavelengths of PAH emission lines from Herbig Ae/Be (HAe/Be) disks
are observed to trend with the effective temperatures of their host
stars \citep{Sloan:2005p6512,Keller:2008p6516}. 
This correlation indicates that PAHs in disks are not merely PAHs from
the diffuse ISM transported unadulterated into circumstellar
environments. Rather, PAHs in disks have been photo-processed, their
chemical bonds altered by radiation from host stars. Possibly
PAHs are continuously created and destroyed by local processes,
e.g., sublimation of grain mantles and photodestruction
\citep{Keller:2008p6516}.
In this paper we do not account explicitly for such processes, i.e., we do not
attempt to calculate the abundance of PAHs from first principles.
Rather we fix the abundance of PAHs using observations, as detailed
in the remainder of this subsection.

Emission from PAHs is detected in an order-unity fraction of HAe/Be stars,
but is rarely seen in T Tauri stars (e.g., \citealt{Geers:2006p3349}, hereafter
G06).  In principle this could mean that PAHs are less abundant in T
Tauri disks, but the more likely explanation is that this is an
observational selection effect: Herbig Ae/Be stars are more luminous
in the ultraviolet (UV) and therefore cause their associated PAHs to
fluoresce more strongly (see, e.g., Figure 9 of G06, which shows how
the PAH intensity drops below the {\it Spitzer} detection threshold
with decreasing stellar effective temperature). Another clue that PAHs
are just as abundant in T Tauri disks as in HAe/Be disks is that those
few T Tauri stars with positive PAH detections tend to have unusually
low mid-infrared continua, allowing PAH emission lines to stand out
more clearly (G06). In other words, those T Tauri systems where PAHs
have been detected are transitional systems, and their PAH abundances
seem no different than in their HAe/Be counterparts.

\citet{Geers:2006p3349} used radiative transfer models to fit the intensities of the 11.2 $\mu$m PAH
fluorescence line in eight Herbig Ae
and T Tauri disks, concluding that the PAH abundance is
$10^{-7}$--$10^{-8}$ per H$_2$ (see their Figure 9).  This
result is highly model dependent. Perhaps the chief source of
uncertainty lies in the grain opacity. Inferred PAH abundances
relative to gas are
sensitive to assumptions about the local grain size distribution and
dust-to-gas ratio because the soft ultraviolet radiation
($\sim$1000--$3000 \,\AA$
wavelength) which causes
PAHs to fluoresce is also absorbed by ambient grains.  Thus the
intensity of PAH emission depends on how many grains are competing
with PAHs for the same illuminating photons. The grain opacity, in
turn, decreases by orders of magnitude as dust settles
(Section \ref{sec_grains}). Because G06 did not account
for dust sedimentation and instead assumed the dust-to-gas ratio in
disks was similar to that of the well-mixed ISM, the PAH abundances
relative to gas that they computed are actually upper limits.  Surface
layer grains have likely settled toward the midplane into regions of
high gas density, and were the local PAH abundance to remain as
inferred at $10^{-7}$--$10^{-8}$ per H$_2$, the ratio of PAH line
intensity to dust continuum would be larger than observed
\citep{Dullemond:2007p3814}.

In our model the number of PAHs per H$_2$ is
\beq{PAH_abundance}
x_{\rm PAH} \equiv \epsilon_{\rm PAH} \times 10^{-6} \, ,
\eeq
where $\epsilon_{\rm PAH}<1$
measures how depleted PAHs are in disk surface layers
relative to PAHs in the diffuse ISM \citep{Li:2001p2937}.
Based on the considerations above, we should combine
the G06 depletion factor of
0.01--0.1 with the grain depletion factor $\epsilon_{\rm grain} \sim 0.001$--0.1,
which parameterizes the reduction of dust opacity due to dust sedimentation.
We thus estimate that $10^{-5} \lesssim \epsilon_{\rm PAH} \lesssim 10^{-2}$
(Table \ref{table_parameters}). In our calculations we select
the parameter combinations $(\epsilon_{\rm grain},\epsilon_{\rm PAH}) = (10^{-1},10^{-2})$
and $(\epsilon_{\rm grain},\epsilon_{\rm PAH}) = (10^{-3},10^{-5})$
which bracket the range of possibilities.

A final point to consider is whether the observed PAHs 
are present at the same
column depths that are relevant for X-ray driven MRI.  The
X-ray stopping column should be compared with the column that presents
optical depth unity to the soft UV radiation driving PAH emission.  In the
model of G06 in which dust has not settled,
photons at wavelengths of 1000--3000 $\AA$ are stopped by
submicron-sized silicate/carbonaceous grains within a hydrogen column
of $\sim$0.005 g cm$^{-2}$ (V.~Geers, private communication, 2010; see also
\citealt{Habart:2004p4185}
 who used similar dust opacities). After we account
for grain settling ($\epsilon_{\rm grain}$), the UV absorption column
increases to $\sim$0.05--5 g cm$^{-2}$. Although model-dependent,
our estimate of the UV absorption column corresponds well
to X-ray stopping columns, and thus to columns that might possibly be
MRI-active.

\subsection{Collisional Charging of PAHs and Grains} 
\label{sec_cc}

Grains and PAHs are modeled as conducting spheres for simplicity.
Electrons and ions collide with and stick to grains and PAHs, charging
them.  When the total electron capture rate by grains and PAHs matches the
total ion capture rate, the distribution of charges carried by
PAHs and grains reaches dynamical equilibrium.  The average
charge state on a PAH/grain $\langle Z \rangle < 0$ because in thermal equilibrium
electrons move more quickly than ions.  Convenient and readily derived
approximations for $\langle Z \rangle$ in various limits
were given by \citet[][hereafter DS]{\DS}. We will find that
$\langle Z \rangle$ ranges between $-1$ and 0 for our PAHs of radius
$6 \AA$, while for our micron-sized grains $\langle Z \rangle \approx
-22$. The remainder of this subsection details how we compute the electron
and ion capture rates.

The rates at which ions or electrons collide with PAHs or grains are
enhanced by Coulomb focusing between static charges, as well as by the
induced dipole force \citep{Natanson:1960p538,Robertson:2008p6609}.
The cross sections can be derived from kinetic theory by considering
the potential between a conducting sphere of radius $s$ and charge
$Ze$, located at a distance $r$ from a charge $q$:
\beq{potential}\phi(Z,r)=\frac{qZe}{r}-\frac{q^{2}
  s^{3}}{2r^{2}(r^{2}-s^{2})} \eeq \citep[e.g.,][]{Jackson:1975p3784}.
The first term is the usual monopole interaction, while the second
arises from the induced dipole (image charges).  For a neutral sphere, the
velocity-dependent cross section derives from applying conservation of
energy and momentum to the second term of (\ref{potential}).
Multiplying this cross section by either the electron or ion velocity, and
averaging over a Maxwellian speed distribution at temperature $T$,
yields the rate coefficient (units of cm$^3$ s$^{-1}$) \beq{nuzero}\alpha =\pi s^{2} S c \left(
  1+ \sqrt{\frac{\pi q^{2}}{2skT}} \right) \, \, \, \mathrm{for} \,
Ze/q=0\eeq where $k$ is the Boltzmann constant, $c=\sqrt{8kT/\pi
  m}$ is the mean speed for either electrons of mass $m=m_{\rm e}$ or
ions of mass $m=m_{\rm X^+}$, and $S$ is the probability that the electron/ion
sticks to the PAH/grain. We will discuss the sticking coefficient $S$ shortly. 

For a charged sphere, the cross
section is enhanced by both terms in (\ref{potential}). There is no
analytical solution for this case, but DS provided the following approximate
formulae: \beq{nu1} \alpha=\pi s^{2} S c [1 -Ze/(q\tau)] (1
+\sqrt{2/(\tau- 2Ze/q)}) \, \, \, \mathrm{for} \, Ze/q<0 \eeq
\beq{nu2} \alpha=\pi s^{2} S c [1 + (4 \tau + 3Ze/q)^{-1/2}]^{2}
\exp(-\beta/\tau) \, \, \, \mathrm{for} \, Ze/q>0\eeq where $\tau
\equiv skT/q^{2}$,
\beq{beta} \beta \equiv \frac{Ze}{qg}-\frac{1}{2g^{2}(g^{2}-1)} \,, \eeq
and $g$ is the solution to the transcendental equation
\beq{g} \frac{2g^{2}-1}{g(g^{2}-1)^{2}}=\frac{Ze}{q} \,. \eeq

Upon colliding with a PAH or grain, the electron or ion sticks with
probability $S$.  For ions, we set $S = S_{\rm{X}^+} = 1$ (DS; IN;
BG).  For electrons colliding with PAHs, $S = S_{\rm{e}}$
depends on the detailed molecular structure of the PAH.
\citet{Allamandola:1989p4607} calculated how
the electron sticking
coefficient increases with both the number of carbon atoms 
and the electron affinity. The dependence on electron
affinity is especially strong. A PAH having
$N_{\rm C} = 32$ and an electron affinity of 0.7 eV
has $S_{\rm e} \approx 3 \times 10^{-5}$ (see their Figure 25),
while the same-sized PAH with an electron affinity of 1 eV
has $S_{\rm e} \approx 10^{-2}$ (see page 769 of their paper).
Estimated electron affinities of real $N_{\rm C}=32$ PAHs (e.g., ovalene
and hexabenzocoronene) exceed 1 eV. 
\citet{Allamandola:1989p4607} stated that ``only for pericondensed PAHs [which
are more stable than catacondensed PAHs] containing considerably more
than 20 C atoms will the electron sticking coefficient approach unity.''
Based on these considerations, we take
$S_{\rm e} = 0.1$ for our $N_{\rm C} = 100$ PAHs.
For the much larger grains we set
$S_{\rm{e}}=1$. 

In our code, the range of charges a grain can possess extends from $Z
= -200$ to +200. We have verified that this range is large enough to
accommodate the entire equilibrium charge distribution,
which for our $\mu$m-sized grains peaks at $-22$ (see Figure
\ref{fig_dist_grains}). Accounting only for a few charges---up to
$|Z|=3$ as did
\citet{Sano:2000p2094}, IN, BG, and TCS---is not necessarily adequate for
micron-sized grains which have fairly large capacitances.  For PAHs we
consider charges $Z$ between $-16$ and +16. Most PAHs will turn out to
have either $Z = 0$ or $-1$. 
Because of their smaller size, a single PAH will be less charged than
a single grain; electrons collide less frequently with a negatively
charged sphere as the radius of the sphere decreases and the Coulomb
potential steepens.

We neglect adsorption of neutral gas-phase species onto grain
surfaces, and any mass increase of grains and PAHs from collisions
with ions. Grain-grain and PAH-grain collisions are negligible and
ignored.  We do account for the possibility that a PAH with a single
negative charge can neutralize by colliding with a PAH with a single
positive charge (reaction 14 in Table \ref{table_eq}), though in
practice this reaction is not significant.

\subsection{Numerical Method of Solution} 
\label{sec_numerical}

The time-dependent rate equations for the abundances of species are
readily constructed from the reactions listed in Table \ref{table_eq}. 
For example, the number density of electrons $n_{\rm e}$ obeys

\begin{eqnarray}
\frac{dn_{\rm e}}{dt}&=&n_{\rm H_2} \zeta -n_{\rm e} \sum_{\rm X^+} \alpha_{\rm X^+,e} \, n_{\rm X^+}  \nonumber \\ 
& & - n_{\rm e} \sum_{Z=-16}^{16} \alpha_{\mathrm{PAH,e}} \, n_{\mathrm{PAH}} 
\nonumber \\
& & - n_{\rm e} \sum_{Z=-200}^{200} \alpha_{\mathrm{grain}, \rm e} \, n_{\mathrm{grain}}
\label{equilibrium_electrons}
\end{eqnarray}
\noindent where the index X$^+$ runs over reactions 7, 8, and 9 in
Table \ref{table_eq}. The first sum over $Z$ occurs over the charge
states of PAHs, while the second sum occurs over the charge states of
grains, with rate coefficients $\alpha$ given in Section \ref{sec_cc}.

The charge distributions of PAHs and grains are governed by recurrence
equations \citep{Parthasarathy:1976p1908, Whitten:2007p2666}, e.g., for PAHs:
\begin{eqnarray}
\frac{d n_{\mathrm{PAH},Z}}{dt} & = & \left( n_{\mathrm{PAH}} \,
  \alpha_{\mathrm{PAH,e}} \right)_{Z+1} n_{\mathrm e} \nonumber \\ 
 & &+ \sum_{\mathrm X^+} \left( n_{\mathrm{PAH}} \, \alpha_{\mathrm
     {PAH,X^+}} \right)_{Z-1} n_{\mathrm X^+}  \nonumber \\
& & - \left( n_{\mathrm{PAH}} \, \alpha_{\mathrm{PAH,e}} \right)_Z n_{\mathrm e}   \nonumber \\ 
& & - \sum_{\mathrm X^+} \left( n_{\mathrm{PAH}} \, \alpha_{\mathrm{PAH,X^+}} \right)_Z n_{\mathrm{X^+}} \,.
\label{equilibrium_pahs}
\end{eqnarray}
\noindent The right-hand side of Equation (\ref{equilibrium_pahs}) 
accounts for all the ways in which PAHs of charge $Z$ can be created
or destroyed by collisions with electrons and ions (reactions 10 and 11
in Table \ref{table_eq}).  When $Z = \pm 1$, Equation
(\ref{equilibrium_pahs}) is supplemented by an extra loss term
accounting for reaction 14.

All rate equations are discretized to first order and advanced
simultaneously using a forward Euler algorithm with a fixed timestep
$\Delta t \leq $ 1\e{-3} s. 
At $t = 0$, all PAHs and grains have $Z=0$ and all hydrogen is in the
form of H$_2$.  In principle we could simply advance the network
forward until the system equilibrates, i.e., until the time rates of
change of the abundances fall below some specified tolerance.  However
the reaction rates in our network span almost 5 orders of
magnitude. Thus, our equations are stiff and a brute-force integration
would require an inordinate number of timesteps.  Metals are typically the
slowest constituent to reach equilibrium because they react with electrons
only slowly by radiative recombination (reaction 9).

To circumvent the bottleneck posed by metals, we proceed as follows.  We run $R$
versions of the code having $R$ evenly spaced initial abundances for
charged metals $n_{\rm M^+}$. For
each run, we initially set $n_{\rm e} = n_{\rm M^+}$ to ensure charge
neutrality. We run each code until the abundances of all species
drift only because of slow changes in $n_{\rm M^+}$. 
We evaluate $dn_{\rm M^+}/dt$ at the end of each run.  The equilibrium
value of $n_{\rm M^+}$ is bracketed by the two runs having opposing
signs for $dn_{\rm M^+}/dt$.  We then start a new iteration with
$R$ runs
having initial metal abundances evenly spaced between the two bounding
runs of the previous iteration. In this way we refine our initial guesses for
$n_{\rm M^+}$ until we arrive at two sets of initial conditions that
differ by less than 30\%. The equilibrium value of $n_{\rm M^+}$ we
report lies at the intersection of the two curves for $n_{\rm
  M^+}(t)$, linearly extrapolated forward in time. Other variables
($n_{\rm{e}}, n_{\rm{HCO^{+}}}$, and $n_{\rm{H_{3}^{+}}}$) are also
extrapolated. The number of runs $R$ at each iteration varies from 2
to 5.   

We use the time $t_{\rm{eq}}$ at which the two extrapolated curves for
$n_{\rm M^+}$ intersect as an estimator of the equilibration time of
the chemical network. For $t_{\rm{eq}}$ to be a robust estimator, it
should be independent of initial conditions. We found that the value
of $t_{\rm{eq}}$ remained constant to within a factor of 3 when
initial conditions varied over 2 orders of magnitude.  Our values
for $t_{\rm eq}$ will be compared to dynamical timescales
$\Omega^{-1}$ in Section \ref{sec_timescale}.

Errors are estimated by monitoring conservation of charge and
conservation of the total number density of PAHs + grains.  Over
$10^{10}$ timesteps, the charge remains constant (at zero) to better
than one part in $10^8$, with similar results for the number density
of PAHs + grains.  As a test of our code, we reproduced the normalized
charge distribution on PAHs computed by \citet[][see their Figure
2a]{Jensen:1991p1393}.

\section{RESULTS} 
\label{sec_results}
In Section \ref{sec_charge_dist}, we describe how
charges distribute themselves on PAHs and grains in dynamical
equilibrium. In Section \ref{sec_ionization_fraction}, we explore how the
free electron and ion abundances vary with increasing PAH
abundance. In Section \ref{sec_AmRe}, we show what all this implies for
the degree of magnetic coupling in disk surface layers, interpreting
our numerical results whenever possible with simple analytic estimates.
In that section we also compute timescales for the chemical network
to equilibrate, and compare to the dynamical timescales over which the MRI
may act. In Section \ref{compare}, we test the validity of our simple
network/code by seeing how closely we can reproduce the results
of more complex networks/codes by BG and TCS.

\subsection{Charge Distributions on PAHs and Grains} 
\label{sec_charge_dist}

\begin{figure}[h!]
\epsscale{1.0}
\plotone{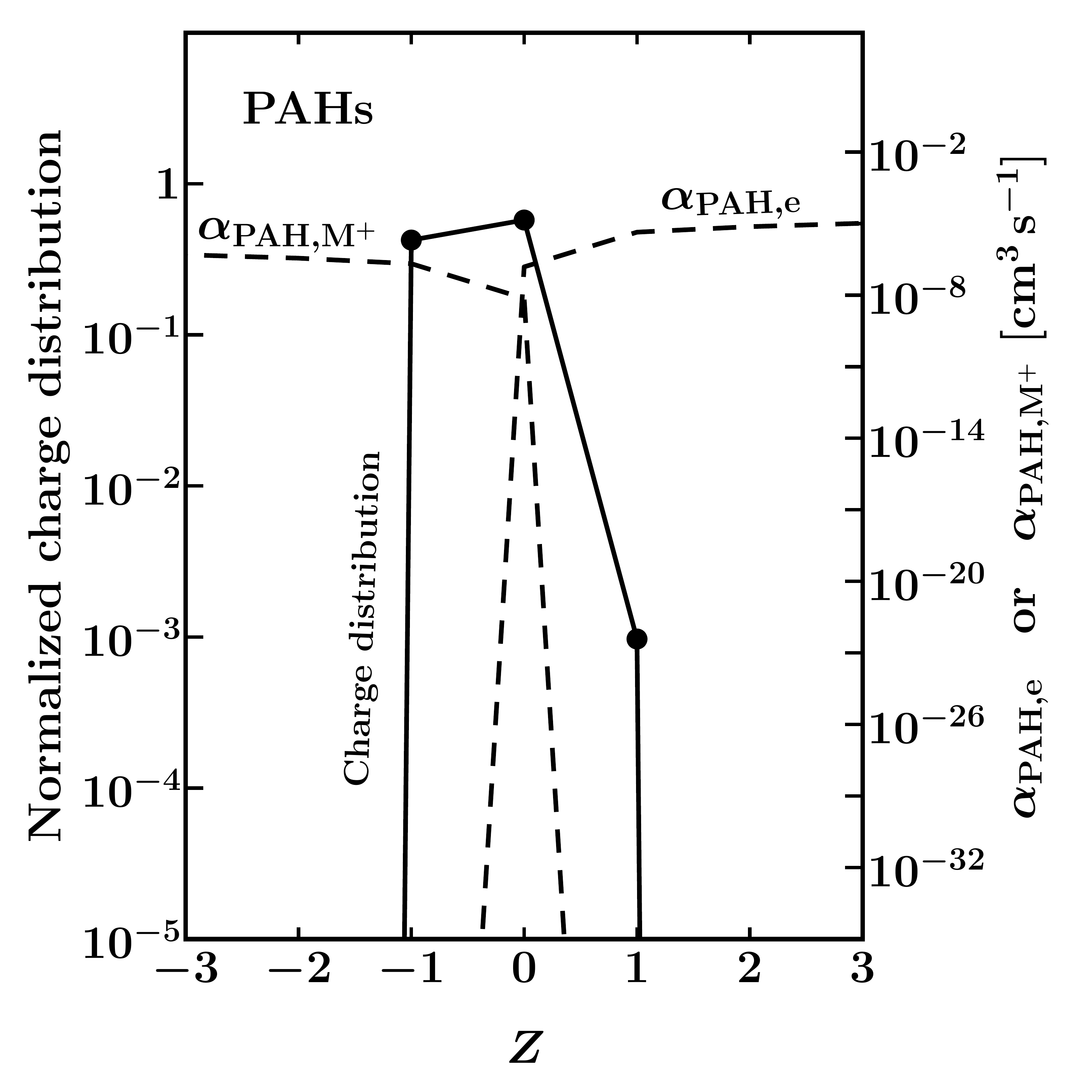}
\caption{Equilibrium charge distribution on PAHs (solid circles, left axis)
for $a = 3$ AU, $\Sigma = 0.3 \, {\rm g \, cm}^{-2}$, $x_{\rm M} =
10^{-8}$ (standard metal abundance), $\epsilon_{\rm PAH} = 10^{-5}$
(low PAH abundance), and $\epsilon_{\rm grain} = 10^{-3}$ (low grain abundance).
The distribution peaks at $Z=0$, approximately 
where the attachment coefficients (dashed lines, right axis)
for electrons with PAHs and metal ions with PAHs cross.
}
\label{fig_dist_pahs}
\end{figure}

\begin{figure}[h!]
\epsscale{1.0}
\plotone{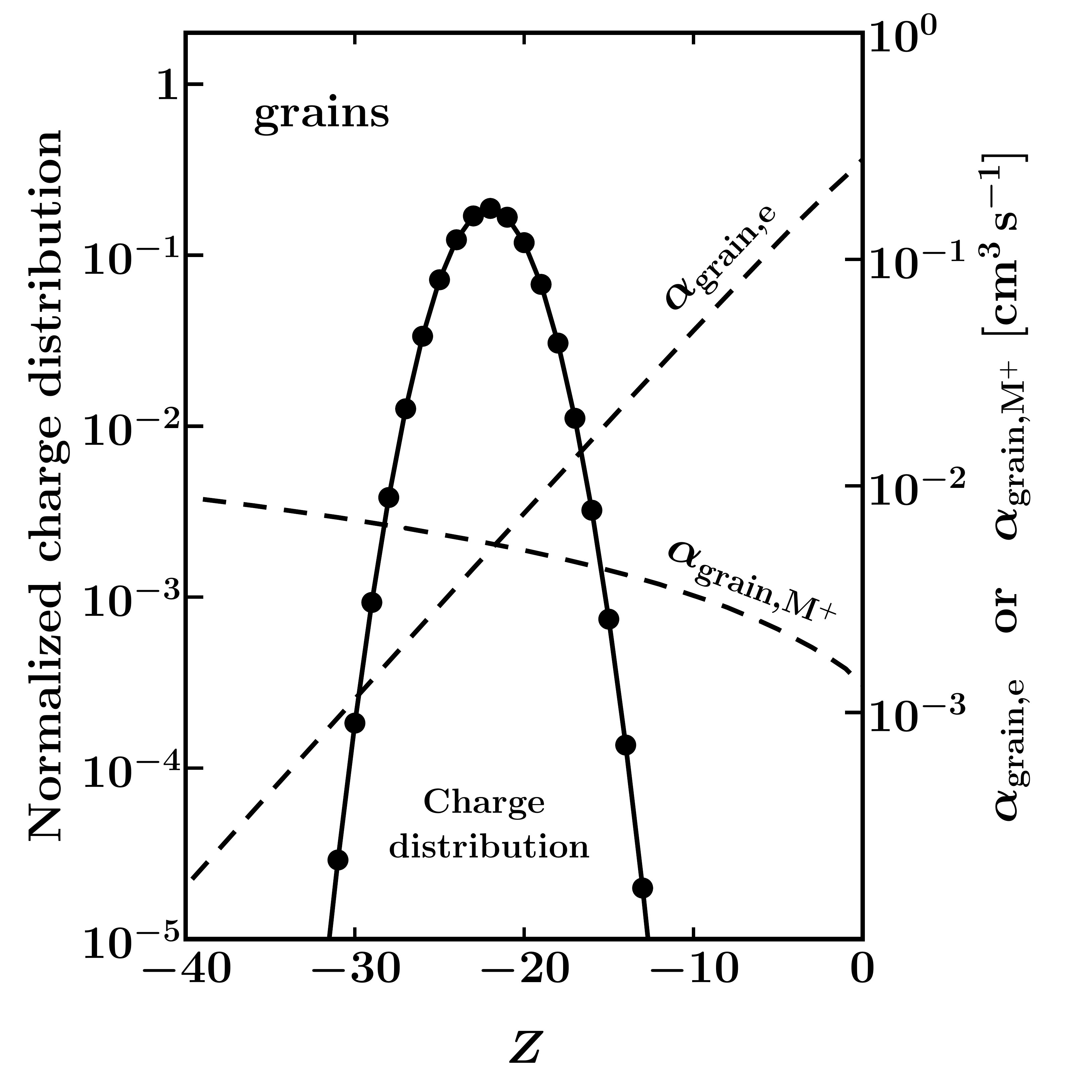}
\caption{Same as Figure \ref{fig_dist_pahs} but for grains.
}
\label{fig_dist_grains}
\end{figure}

Figures \ref{fig_dist_pahs} and \ref{fig_dist_grains} show the
charge distributions on PAHs and grains, respectively, for the case $a
= 3$ AU, $\Sigma = 0.3 \, {\rm g\, cm}^{-2}$, $x_{\rm M} =
10^{-8}$ (standard metal abundance), $\epsilon_{\rm PAH} = 10^{-5}$
(low PAH abundance), and $\epsilon_{\rm grain} = 10^{-3}$ (low grain
abundance). Most of the PAHs either have $Z_{\rm PAH} = 0$ or $Z_{\rm
  PAH} = -1$.  For grains, the average charge state (the peak of the
distribution) is $\langle Z_{\rm grain} \rangle \approx -22$. The
shape of the charge distribution for grains approaches the Gaussian
given by Equation (4.15) of DS.

We may understand $\langle Z \rangle$ simply. Consider the PAHs;
identical considerations apply to grains. We take the limit that the
dominant ions are metals and the limit that the total charge carried
by PAHs is much less than the free charge. Together these limits imply
that $x_{\rm e} \approx x_{\rm M^+}$. Then detailed balance between
forward and reverse reaction rates dictates that (cf. Equation
\ref{equilibrium_pahs}): \beq{pah_balance} \left( n_{\mathrm{PAH}} \,
  \alpha_{\mathrm{PAH,e}} \right)_{Z+1}=\left( n_{\mathrm{PAH}} \,
  \alpha_{\mathrm{PAH,M^+}} \right)_Z \,\, .  \eeq From this equation
it is evident that if ever the rate coefficients
$(\alpha_{\mathrm{PAH,e}})_{Z+1}$ and $(\alpha_{\mathrm{PAH,M^+}})_Z$
were to be equal, the densities $(n_{\mathrm{PAH}})_{Z+1}$ and
$(n_{\mathrm{PAH}})_Z$ would be equal, i.e., the charge distribution
would be at an extremum. Thus we may estimate the average charge
$\langle Z \rangle$ by merely plotting the rate coefficients
$\alpha_{\mathrm{PAH,M^+}}$ and $\alpha_{\mathrm{PAH,e}}$ against $Z$
and seeing where the curves intersect. This exercise is performed in
Figures \ref{fig_dist_pahs} and \ref{fig_dist_grains}. Indeed what the
full numerical model gives for $\langle Z \rangle$ is close to the $Z$
for which the curves for the rate coefficients intersect.
(Of course, perfect agreement cannot be obtained because it is never
strictly true that $(\alpha_{\mathrm{PAH,e}})_{Z+1} =
(\alpha_{\mathrm{PAH,M^+}})_Z$.)

\begin{figure}[h!]
\epsscale{1.0}
\plotone{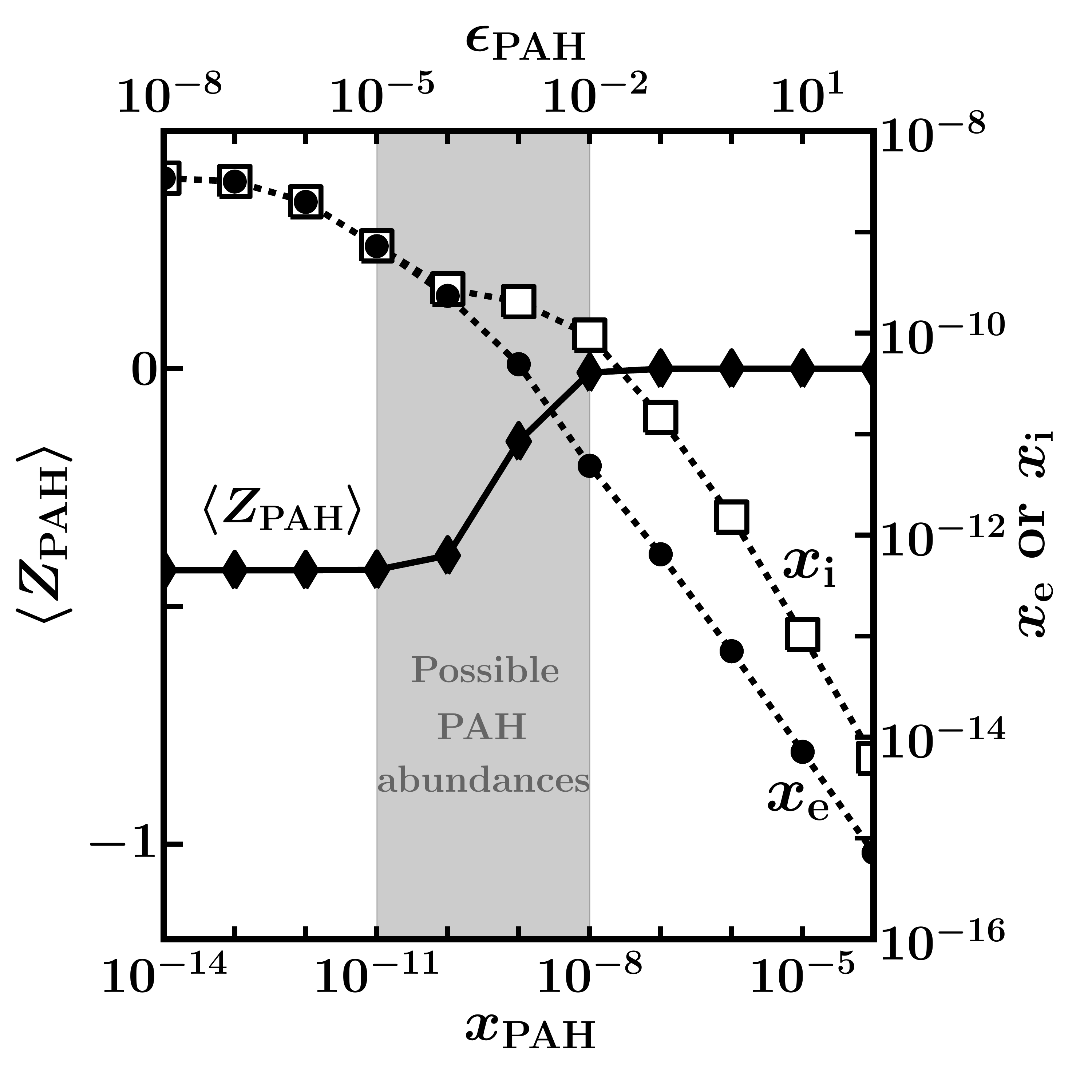}
\caption{Average charge state of PAHs as
  a function of PAH abundance (solid diamonds, left axis). Dashed
  lines show simulation results for fractional electron abundance $x_{\rm{e}}$
  (solid circles, right axis) and fractional ion abundance $x_{\rm{i}}$ (open
  squares, right axis). The shaded region marks observationally inferred PAH
  abundances, measured by number either
  relative to H$_2$ ($x_{\rm PAH}$, bottom axis) or relative to the
  PAH abundance in the diffuse ISM (depletion factor $\epsilon_{\rm PAH}
  \equiv x_{\rm PAH}/10^{-6}$, top axis).  Parameters for this run are $a=3$ AU,
  $x_{\rm{M}}=10^{-8}$, $\Sigma = 0.3$ g cm$^{-2}$, and
  $\epsilon_{\rm{grain}}=0$ (kept at zero to
  isolate the effect of PAHs). The shift to $\langle
  Z_{\rm{PAH}}\rangle= 0$ occurs when there are so many PAHs
  that they begin to adsorb most of the free charge. At this point
  $x_{\rm i}$ and $x_{\rm e}$ diverge; see also Figure \ref{fig_x_pah_crit_M8}.
}
\label{fig_pahs_Z}
\end{figure}

\begin{figure}[h!]
\epsscale{1.0}
\plotone{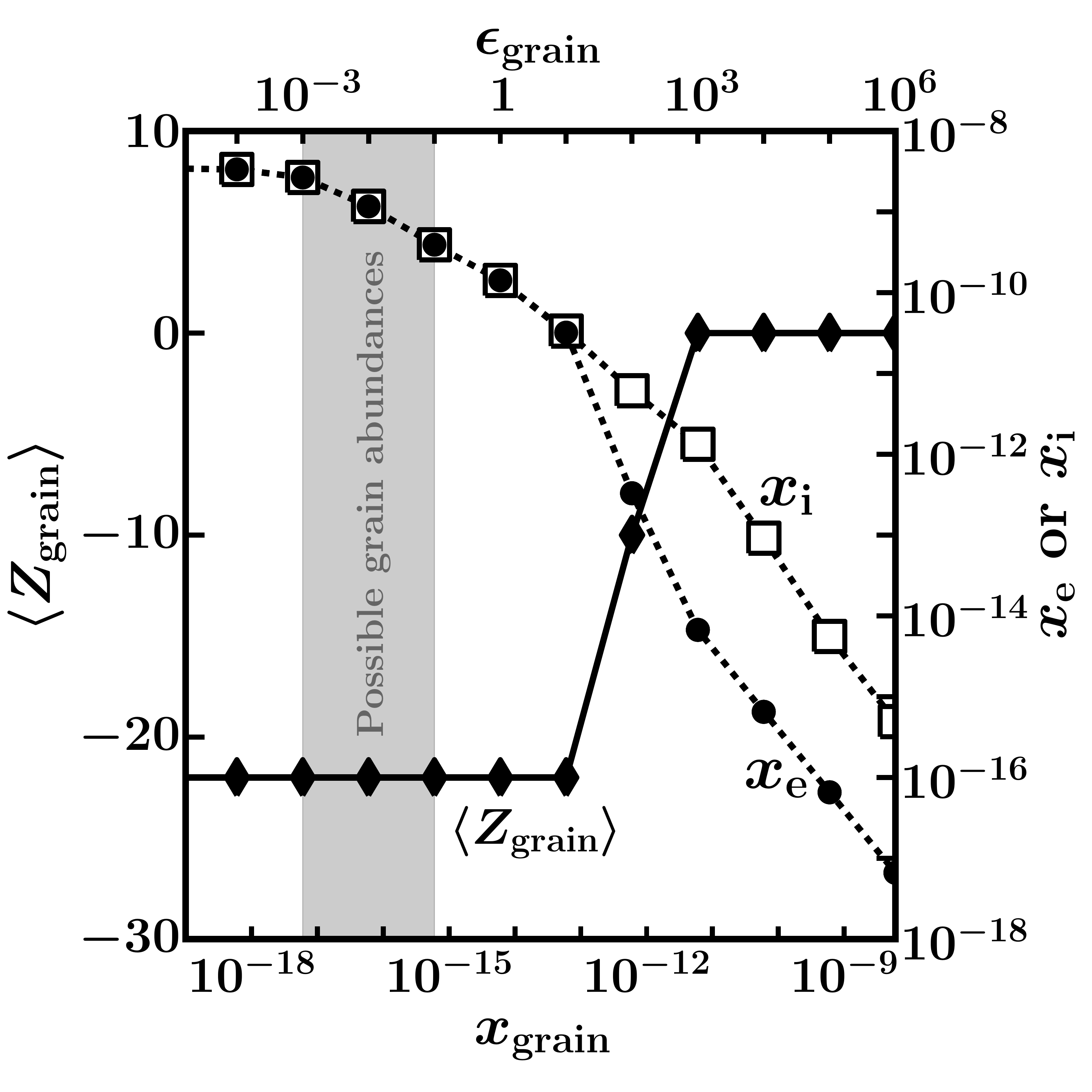}
\caption{Same as Figure \ref{fig_pahs_Z} but for grains.  Parameters
  for this run are $a=3$ AU, $x_{\rm{M}}=10^{-8}$, $\Sigma = 0.3$
  g cm$^{-2}$, and $\epsilon_{\rm{PAH}}=0$ (kept at zero to isolate the
  effect of grains). Our grains all have radii of $1 \, \mu$m, which
  is so large that their corresponding abundance as inferred from
  observation (shaded region) is too low to significantly affect the
  amount of free charge.  }
\label{fig_grains_Z}
\end{figure}

There is another, even simpler limit where $\langle Z \rangle$ may be
estimated. In the extreme case that the gas is so saturated with
grains or PAHs that practically no free charges are left, we must have
$\langle Z \rangle \rightarrow 0$.  Figure \ref{fig_pahs_Z} shows the
results of an experiment using our full code in which we increase
$\epsilon_{\rm PAH}$ until this regime is reached.  For this Figure, the
grain abundance is set to zero to isolate the effects of PAHs. Figure
\ref{fig_grains_Z} is analogous; $\epsilon_{\rm grain}$ is increased
while the PAH abundance is held fixed at zero. Both figures follow
the transition from $\langle Z \rangle \neq 0$ to $\langle Z \rangle
\rightarrow 0$.  Observationally inferred values for $\epsilon_{\rm
  grain}$ (see the shaded region of Figure \ref{fig_grains_Z}) are
never so high as to cross into the $\langle Z_{\rm grain} \rangle
\rightarrow 0$ regime.  By contrast, Figure \ref{fig_pahs_Z} shows
that PAHs may be sufficiently abundant in disks 
that they impact the density of free charges. The critical PAH
abundance $x_{\rm PAH}^\star$ dividing the $\langle Z_{\rm PAH}
\rangle \neq 0$ limit from the $\langle Z_{\rm PAH} \rangle
\rightarrow 0$ limit is the one for which an electron attaches itself
to a PAH as frequently as it recombines with an ion. This critical
abundance is discussed further in Section \ref{sec_ionization_fraction}.

\subsection{Ionization Fraction vs.~PAH Abundance}
\label{sec_ionization_fraction}

Figure \ref{fig_x_pah_crit_M8} plots the fractional electron and ion
densities, $x_{\rm
  e}$ and $x_{\rm i}$, against the PAH abundance $x_{\rm
  PAH}$, for $a = 3$ AU, $\Sigma = 0.3 $ g cm$^{-2}$, $x_{\rm M} = 10^{-8}$
(standard metal abundance), and $\epsilon_{\rm grain}= 10^{-3}$--$10^{-1}$ (see
the figure caption for how $\epsilon_{\rm grain}$ is
assigned to each $\epsilon_{\rm PAH}$).
Figure \ref{fig_x_pah_crit_M6} is identical except that
it considers the metal-rich case $x_{\rm M} = 10^{-6}$.
The primary ions in both cases are atomic metals and
HCO$^+$ molecules. At low PAH abundances, charged metal ions are the
most abundant. As the number of PAHs is increased, HCO$^+$ becomes the
dominant ion. See Table \ref{table_abundances} for a precise breakdown of component
ion densities for our standard metal abundance case.

According to Figures \ref{fig_x_pah_crit_M8} and
\ref{fig_x_pah_crit_M6}, the electron and ion densities are nearly
equal and constant with $x_{\rm PAH}$ as long as $x_{\rm PAH}$ is not
too large. In going from the standard metal abundance of $x_{\rm M} =
10^{-8}$ to the metal-rich case of $x_{\rm M} = 10^{-6}$, the free
charge abundance increases by an order of magnitude. Once $x_{\rm
  PAH}$ exceeds some critical abundance $x_{\rm PAH}^\star$, the
electron and ion densities diverge---the ion density is higher, and
the balance of negative charges is carried by PAHs.  In the limit
$x_{\rm PAH} \gg x_{\rm PAH}^\star$, both the electron and ion
densities decrease with increasing PAH abundance in an approximately
inverse linear way.  All of this behavior can be understood
analytically as follows.

\begin{deluxetable*}{llllllllllll}
\tabletypesize{\tiny}
\tablewidth{0pt}
\tablecaption{Densities of Charged Species for our Standard Model
  $(x_{\rm{M}}=10^{-8}, L_{\rm X} = 10^{29} \, {\rm erg} \, {\rm s}^{-1})$\label{table_abundances}}
\tablehead{\colhead{$ $} &\colhead{$ $} &\colhead{$ $} &\colhead{$ $} &\colhead{$ $} &\colhead{$ $} &\colhead{$ $} &\colhead{$ $} &\colhead{$ $} &\colhead{$ $} &\colhead{$ $} &\colhead{$ $} }

\startdata
\cutinhead{\boldmath $a=3 \, \mathrm{AU},\,\, \epsilon_{\mathrm{PAH}}=10^{-5},\,\, \epsilon_{\mathrm{grain}}=10^{-3}$}
$\Sigma$  & $n_{\mathrm{H_2}}$  & $n_{\rm{e}}$ & $n_{\rm{M^{+}}}$ &
$n_{\mathrm{HCO^+}}$ & $n_{\mathrm{H_3^+}}$ &  $x_{\mathrm{e}}$&
$x_{\mathrm{i}}$ &  $x_{\mathrm{PAH}}$&  $x_{\mathrm{grain}}$&$\langle Z_{\rm{PAH}} \rangle$  & $\langle Z_{\rm{grain}} \rangle$ \\

6\e{-3} & 1\e{9} & 2\e{1} & 1\e{1} & 8\e{0} & 5\e{-1}&2\e{-8}&2\e{-8}&1\e{-11} &6\e{-18} &$-$4\e{-1}&$-22$\\
4\e{-2} & 7\e{9} & 4\e{1} & 4\e{1} & 4\e{0} & 8\e{-2}&6\e{-9}&6\e{-9}&1\e{-11} &6\e{-18} &$-$4\e{-1}&$-22$\\
3\e{-1} &5\e{10} &3\e{1} & 3\e{1} & 1\e{0} & 3\e{-3}&6\e{-10}&6\e{-10}&1\e{-11} &6\e{-18} &$-$4\e{-1}&$-22$\\
2\e{0} & 4\e{11} & 1\e{1} & 1\e{1} & 4\e{-1} & 1\e{-4}&4\e{-11}&4\e{-11}&1\e{-11} &6\e{-18} &$-$4\e{-1}&$-22$\\
1\e{1} & 2\e{12} & 6\e{-1} & 3\e{0} & 3\e{-2} & 6\e{-6}&2\e{-13}&1\e{-12}&1\e{-11} &6\e{-18}  &$-$1\e{-1}&$-15$\\
6\e{1} & 1\e{13} &7\e{-4} & 5\e{-3} & 2\e{-5} & 3\e{-9}&6\e{-17}&4\e{-16}&1\e{-11} &6\e{-18}  &$-$4\e{-5}&$-14$\\

& & & & & &  &  & &  \\

\cutinhead{\boldmath $a=3 \, \mathrm{AU},\,\, \epsilon_{\mathrm{PAH}}=10^{-2},\,\, \epsilon_{\mathrm{grain}}=10^{-1}$}
$\Sigma$  & $n_{\mathrm{H_2}}$  & $n_{\rm{e}}$ & $n_{\rm{M^{+}}}$ & $n_{\mathrm{HCO^+}}$ & $n_{\mathrm{H_3^+}}$ &  $x_{\mathrm{e}}$& $x_{\mathrm{i}}$&  $x_{\mathrm{PAH}}$&  $x_{\mathrm{grain}}$&$\langle Z_{\rm{PAH}} \rangle$  & $\langle Z_{\rm{grain}} \rangle$ \\

6\e{-3}&1\e{9} & 1\e{1} & 5\e{-1} & 1\e{1} & 5\e{-1}&9\e{-9}&1\e{-8}&1\e{-8} &6\e{-16} &$-$4\e{-1}&$-21$\\
4\e{-2}&7\e{9} & 5\e{0} & 1\e{0} & 2\e{1} & 8\e{-2}&7\e{-10}&2\e{-9}&1\e{-8} &6\e{-16} &$-$2\e{-1}&$-17$\\
3\e{-1}&5\e{10} & 2\e{-1} & 7\e{-1} & 3\e{0} & 3\e{-3}&4\e{-12}&8\e{-11}&1\e{-8} &6\e{-16} &$-$8\e{-3}&$-9$\\
2\e{0}&4\e{11} & 1\e{-2} & 7\e{-2} & 2\e{-1} & 1\e{-4}&4\e{-14}&8\e{-13}&1\e{-8} &6\e{-16} &$-$8\e{-5}&$-9$\\
1\e{1}&2\e{12} & 6\e{-4} & 4\e{-3} & 1\e{-2} & 6\e{-6}&3\e{-16}&6\e{-15}&1\e{-8} &6\e{-16} &$-$1\e{-7}&$-9$\\
6\e{1}&1\e{13} & 3\e{-7} & 2\e{-6} & 5\e{-6} & 3\e{-9}&2\e{-20}&6\e{-19}&1\e{-8} &6\e{-16} &$-$1\e{-10}&$-$3\e{-3}\\

& & & & & &  &  & & \\

\cutinhead{ \boldmath$a=30 \, \mathrm{AU},\,\, \epsilon_{\mathrm{PAH}}=10^{-5},\,\, \epsilon_{\mathrm{grain}}=10^{-3}$}
$\Sigma$  & $n_{\mathrm{H_2}}$  & $n_{\rm{e}}$ & $n_{\rm{M^{+}}}$ &
$n_{\mathrm{HCO^+}}$ & $n_{\mathrm{H_3^+}}$ &  $x_{\mathrm{e}}$&
$x_{\mathrm{i}}$ &  $x_{\mathrm{PAH}}$&  $x_{\mathrm{grain}}$&$\langle Z_{\rm{PAH}} \rangle$  & $\langle Z_{\rm{grain}} \rangle$ \\

6\e{-3} & 6\e{7} & 5\e{-1} & 4\e{-1} & 1\e{-1} & 7\e{-3}&9\e{-9}&9\e{-9}&1\e{-11} &6\e{-18} & $-$3\e{-1}&$-9$\\
4\e{-2} & 4\e{8} & 9\e{-1} & 9\e{-1} & 4\e{-2} & 8\e{-4}&2\e{-9}&2\e{-9}&1\e{-11} &6\e{-18} & $-$3\e{-1}&$-9$\\
3\e{-1} & 3\e{9} & 4\e{-1} & 4\e{-1} & 2\e{-2} & 2\e{-5}&2\e{-10}&2\e{-10}&1\e{-11} &6\e{-18} &$-$3\e{-1}&$-9$\\
2\e{0} & 2\e{10} & 1\e{-1} & 2\e{-1} & 6\e{-3} & 1\e{-6}&8\e{-12}&1\e{-11}&1\e{-11} &6\e{-18} &$-$3\e{-1}&$-9$\\
1\e{1} & 1\e{11} & 7\e{-3} & 5\e{-2} & 4\e{-4} & 7\e{-8}&7\e{-14}&4\e{-13}&1\e{-11} &6\e{-18} &$-$4\e{-2}&$-6$\\
6\e{1} & 6\e{11} & 3\e{-6} & 5\e{-5} & 2\e{-7} & 3\e{-11}&5\e{-18}&1\e{-16}&1\e{-11} &6\e{-18} &$-$8\e{-6}&$-4$\\

& & & & & &  &  & &  \\

\cutinhead{\boldmath $a=30 \, \mathrm{AU},\,\, \epsilon_{\mathrm{PAH}}=10^{-2},\,\, \epsilon_{\mathrm{grain}}=10^{-1}$}
$\Sigma$  & $n_{\mathrm{H_2}}$  & $n_{\rm{e}}$ & $n_{\rm{M^{+}}}$ & $n_{\mathrm{HCO^+}}$ & $n_{\mathrm{H_3^+}}$ &  $x_{\mathrm{e}}$& $x_{\mathrm{i}}$&  $x_{\mathrm{PAH}}$&  $x_{\mathrm{grain}}$&$\langle Z_{\rm{PAH}} \rangle$  & $\langle Z_{\rm{grain}} \rangle$ \\

6\e{-3} & 6\e{7} & 2\e{-1} & 9\e{-3} & 3\e{-1} & 7\e{-3}&3\e{-9}&5\e{-9}&1\e{-8} &6\e{-16} & $-$2\e{-1} & $-8$\\
4\e{-2} & 4\e{8} & 5\e{-2} & 2\e{-2} & 3\e{-1} & 8\e{-4}&1\e{-10}&8\e{-10}&1\e{-8} &6\e{-16} & $-$7\e{-2} & $-6$\\
3\e{-1} & 3\e{9} & 2\e{-3} & 9\e{-3} & 3\e{-2} & 2\e{-5}&7\e{-13}&2\e{-11}&1\e{-8} &6\e{-16} & $-$2\e{-3} & $-4$\\
2\e{0} & 2\e{10} & 1\e{-4} & 8\e{-4} & 2\e{-3} & 1\e{-6}&7\e{-15}&2\e{-13}&1\e{-8} &6\e{-16} & $-$2\e{-5} & $-4$\\
1\e{1} & 1\e{11} & 8\e{-6} & 5\e{-5} & 1\e{-4} & 7\e{-8}&7\e{-17}&2\e{-15}&1\e{-8} &6\e{-16} & $-$1\e{-9} & $-3$\\
6\e{1} & 6\e{11} & 3\e{-9} & 2\e{-8} & 5\e{-8} & 3\e{-11}&5\e{-21}&1\e{-19}&1\e{-8} &6\e{-16} & $-$1\e{-10} & $-$2\e{-3}\\

\enddata
\tablecomments{The surface density $\Sigma$ has units of g cm$^{-2}$;
  number densities $n$ have units of cm$^{-3}$; and fractional
  densities $x$ are measured per H$_2$. The ion density $x_{\rm i} =x_{\rm{M^{+}}} +x_{\rm{HCO^{+}}}$.}
\end{deluxetable*}

\subsubsection{Analytical Model for Ionization Fraction vs.~PAH Abundance} 
\label{sec_analytical}

Our code's results for $x_{\rm e}(x_{\rm PAH})$, $x_{\rm i}(x_{\rm
  PAH})$, and $x_{\rm PAH}^\star$ may be understood using the
following simple model.  The model consists only of X-rays, molecular
hydrogen, electrons, PAHs, and one ion species---either HCO$^+$ for our
standard metal abundance case, or ionized metals $\mathrm M^+$ for the
metal-rich case.  In the simplified model, X-ray ionization of a
hydrogen molecule produces a free electron and---skipping the entire
reaction chain---one ion.  The system reduces to the rate equations
\beq{el_balance} \frac{dn_{\rm e}}{dt} = \zeta n_{\rm H_{2}} - n_{\rm
  e} n_{\rm{i}} \alpha_{\mathrm {i,e}} - n_{\rm e} x_{\mathrm{PAH}}
n_{\rm H_2} (\alpha_{\rm PAH,e})_{\langle Z \rangle} \eeq
\beq{ion_balance} \frac{dn_{\rm i}}{dt} = \zeta n_{\rm H_{2}} -
n_{\rm{i}} n_{\rm e} \alpha_{\rm i,e} - n_{\rm{i}} x_{\mathrm{PAH}}
n_{\rm H_2} (\alpha_{\rm PAH,i})_{\langle Z \rangle}\eeq for the
electron and ion densities, $n_{\rm e}$ and $n_{\rm i}$. The subscript
$\mathrm i$ denotes either HCO$^+$ or $\mathrm M^+$.

In the simplified model, all PAHs with abundance $x_{\mathrm{PAH}}n_{\rm
  H_2}$ are assumed to be identically charged. We set this common
charge equal to the average charge state $\langle Z \rangle$, results
for which were given in Section \ref{sec_charge_dist}.

We exclude our large, micron-sized grains from the analytic
model. Although these grains were useful for inferring PAH abundances
from observations (Section \ref{sec_PAHs}), their collective surface area is
too low to significantly influence the electron chemistry in any of our
model runs. Of course, because PAHs and grains are both modeled the
same way, i.e., as spherical conductors, all of the equations below would
still be valid were we to replace PAHs with grains.

\begin{figure}[h!]
\epsscale{0.9}
\plotone{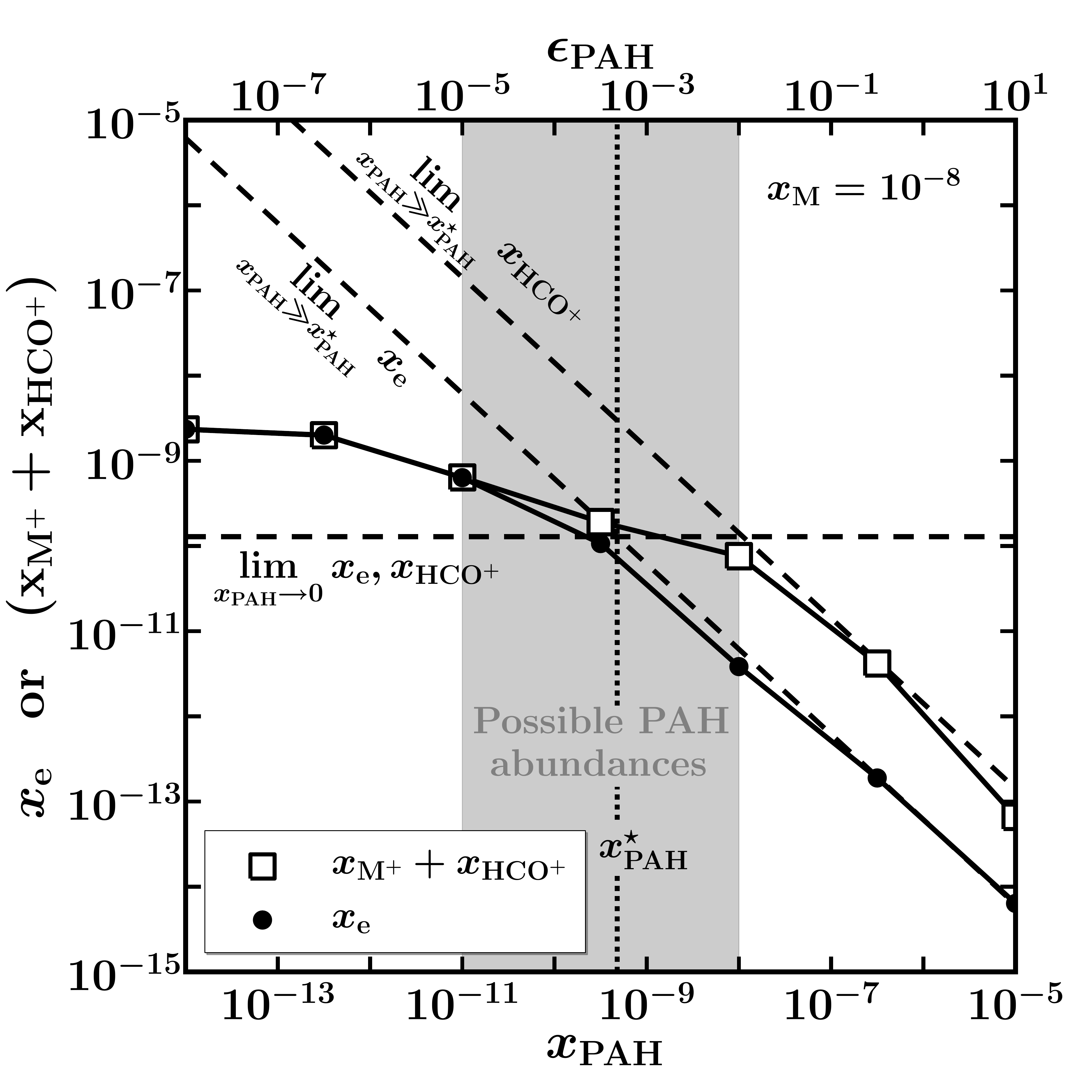}
\caption{Ionization fraction as a function of PAH abundance for
  $x_{\rm{M}}=10^{-8}$ (standard metal abundance), $a=3$ AU, and
  $\Sigma=0.3$ g cm$^{-2}$.  Dashed lines: asymptotic values for
  $x_{\rm{e}}$ and $x_{\rm{HCO^+}}$ of the simplified model of Section 
  \ref{sec_analytical}. Solid lines: simulation results for fractional
  electron abundance $x_{\rm{e}}$ (solid circles) and fractional ion
  abundance $x_{\rm{M^+}}+x_{\rm{HCO^+}}$ (open squares). The dotted
  vertical line marks $x_{\mathrm{PAH}}^{\star}$ (Equation
  \ref{x_pah_crit}), which roughly divides the regime of ``low PAH
  abundance'' where electron and ion densities are equal and
  insensitive to PAH abundance, from the regime of ``high PAH
  abundance'' where the ion density exceeds that of electrons and both
  decrease approximately as $1/x_{\rm PAH}$. The behavior at high PAH
  abundance is independent of the metal abundance; compare with Figure
  \ref{fig_x_pah_crit_M6}. The shaded region marks observationally
  inferred PAH abundances and happens to span the transition from low
  to high PAH regimes. Simulation data use $\epsilon_{\rm grain} =
  10^{-3}$ for $\epsilon_{\rm PAH} \leq 10^{-5}$; $\epsilon_{\rm
    grain} = 10^{-2}$ for $\epsilon_{\rm PAH} = 10^{-3.5}$; and
  $\epsilon_{\rm grain} = 10^{-1}$ for $\epsilon_{\rm PAH} \geq
  10^{-2}$.  }
\label{fig_x_pah_crit_M8}
\end{figure}

\paragraph{Standard metal abundance.}\label{sma}
As stated above, we assume for this case that
all ions are HCO$^+$ molecules and neglect M$^+$.
We may solve for $x_{\rm e}$ and $x_{\rm i} = x_{\rm HCO^+}$
in the limits of low and high PAH abundance.
In the limit of low $x_{\rm PAH}$, the rightmost terms in Equations
(\ref{el_balance}) and (\ref{ion_balance}) can be ignored, yielding in 
steady state:
\begin{eqnarray}
x_{\rm e} =x_{\rm{HCO^{+}}} & = & \sqrt{\zeta /n_{\rm H_{2}}
  \alpha_{\rm{HCO^{+},e}}}  \label{low_pah} \\ 
 & \sim & 10^{-10} \left(\frac{L_{\rm{X}}}{10^{29} \,
    \rm{erg \,s^{-1}}}\right)^{1/2}  \nonumber \\ 
& & \times \left( \frac{\Sigma}{0.3 \,{\rm g\,  cm}^{-2}}
 \right)^{-1/2} \left( \frac{a}{3 \, {\rm AU}}
 \right)^{-0.5} \label{low_pah_a}\\ \nonumber
\\ 
& & {\rm
   \,\, for \,\, standard \,\, metals \,\, and \,\, low \,\, PAHs. } \nonumber
\end{eqnarray}
In going from Equations (\ref{low_pah}) to (\ref{low_pah_a}) we account for
the distance dependence of temperature but assume material is optically
thin to X-rays.  The square-root law of Equation (\ref{low_pah}) is often used by
other workers (e.g., \citealt{Gammie:1996p3339}; \citealt{Glassgold:1997p2130}). It is
plotted as a horizontal dashed line in Figure \ref{fig_x_pah_crit_M8},
and should be compared with the curves for $x_{\rm e}$ and $x_{\rm i}$
from our code, plotted as solid lines. In the limit of low $x_{\rm
  PAH}$, the electron and ion abundances computed from the code are
nearly constant with $x_{\rm PAH}$, as predicted by the analytic
model. However, the results from the code sit above the line for Equation
(\ref{low_pah}) by about an order of magnitude. The factor of 10
offset arises because Equation (\ref{low_pah}) ignores ionized metals,
which recombine with electrons much more slowly than does HCO$^+$ and
which remain abundant compared to HCO$^+$ in our standard model.
The offset also implies that reducing the total metal
abundance below that of our standard model ($x_{\rm M} = 10^{-8}$) can
only decrease $x_{\rm e}$ and $x_{\rm i}$ by at most a factor of
$\sim$10. In this sense our uncertainty in the metal abundance
(Section \ref{sec_metals}) has only a limited impact on the ionization
fraction, assuming $x_{\rm M} < 10^{-8}$. See also
Section \ref{sec_lx} where we consider the case $x_{\rm M}=0$.

In the limit of high PAH abundance, electron recombination on PAHs
dominates electron recombination with HCO$^+$. Low equilibrium
abundances of free electrons imply the average charge on PAHs $\langle
Z \rangle \rightarrow 0$ (Section \ref{sec_charge_dist}). Then the
steady-state solutions to Equations (\ref{el_balance}) and (\ref{ion_balance})
are, respectively,
\begin{subequations}
\begin{eqnarray}
    x_{\rm e}& = & \frac{\zeta}{x_{\mathrm{PAH}} n_{\rm H_2} (\alpha_{\rm PAH,e})_{\langle Z \rangle = 0}} \label{el_high_pah}\\
    x_{\mathrm{HCO}^{+}} & = & \frac{\zeta}{x_{\mathrm{PAH}} n_{\rm
        H_2} (\alpha_{\rm PAH,HCO^+})_{\langle Z \rangle = 0}} \label{ion_high_pah}\\
\nonumber
\\
& &  {\rm for \,\, high \,\, PAHs}. \nonumber
\end{eqnarray}
\end{subequations}
\noindent Note that in this limit of high PAH abundance,
$x_{\rm e} < x_{\rm HCO^+}$ because
$\alpha_{\rm PAH,e} > \alpha_{\rm PAH,HCO^+}$; electrons move faster
than ions. The remaining negative charge required to maintain charge
neutrality is carried by PAHs.  That $x_{\rm e} \neq x_{\rm HCO^+}$ is
relevant for the computation of $Am$ and $Re$ (Section \ref{sec_AmRe})
because $Am$ depends on
$x_{\rm HCO^+}$ (ions carry the bulk of the momentum in a plasma)
while $Re$ depends on $x_{\rm e}$ (electrons are the most mobile
charge carriers). Expressions (\ref{el_high_pah}) and (\ref{ion_high_pah})
are plotted as diagonal dashed lines in Figure \ref{fig_x_pah_crit_M8};
they compare well with the full numerical results, shown
as solid lines, in the limit of high $x_{\rm PAH}$.

The critical PAH abundance dividing these limits is estimated 
by equating Equations (\ref{el_high_pah}) to (\ref{low_pah}):
\begin{eqnarray}
x_{\mathrm{PAH}}^{\star} & = & \sqrt{\frac{\zeta \alpha_{\rm
      HCO^{+},e}}{n_{\rm H_{2}} (\alpha_{\rm PAH,e})^2_{\langle Z
      \rangle = 0}}} \label{x_pah_crit}\\
 & \sim & 5 \times 10^{-10} \left(\frac{L_{\rm{X}}}{10^{29} \,
    \rm{erg \,s^{-1}}}\right)^{1/2} \nonumber \\ \nonumber 
& & \times \left(\frac{\Sigma}{0.3 \,
    \rm{g \, cm^{-2}}}\right)^{-1/2} \left(\frac{a}{3
    \,\rm{AU}}\right)^{-0.2}  \\ \nonumber
\\
& & {\rm for \,\, standard \,\, metals}. \nonumber
\end{eqnarray}
The critical value $x_{\rm PAH}^\star$ marks
the abundance at which PAHs start to reduce
significantly the number of free charges, i.e.,
the abundance at which electron recombination on PAHs
becomes competitive with electron recombination with molecular ions.
It is plotted in Figure \ref{fig_x_pah_crit_M8} as a vertical line and
does reasonably well at delineating the regime where the ionization
fraction does not depend on PAHs from the regime where it
does.  Note that possible PAH abundances as inferred from observations
(Section \ref{sec_PAHs}) happen to straddle $x_{\rm PAH}^\star$.

\begin{figure}[h!]
\epsscale{0.9}
\plotone{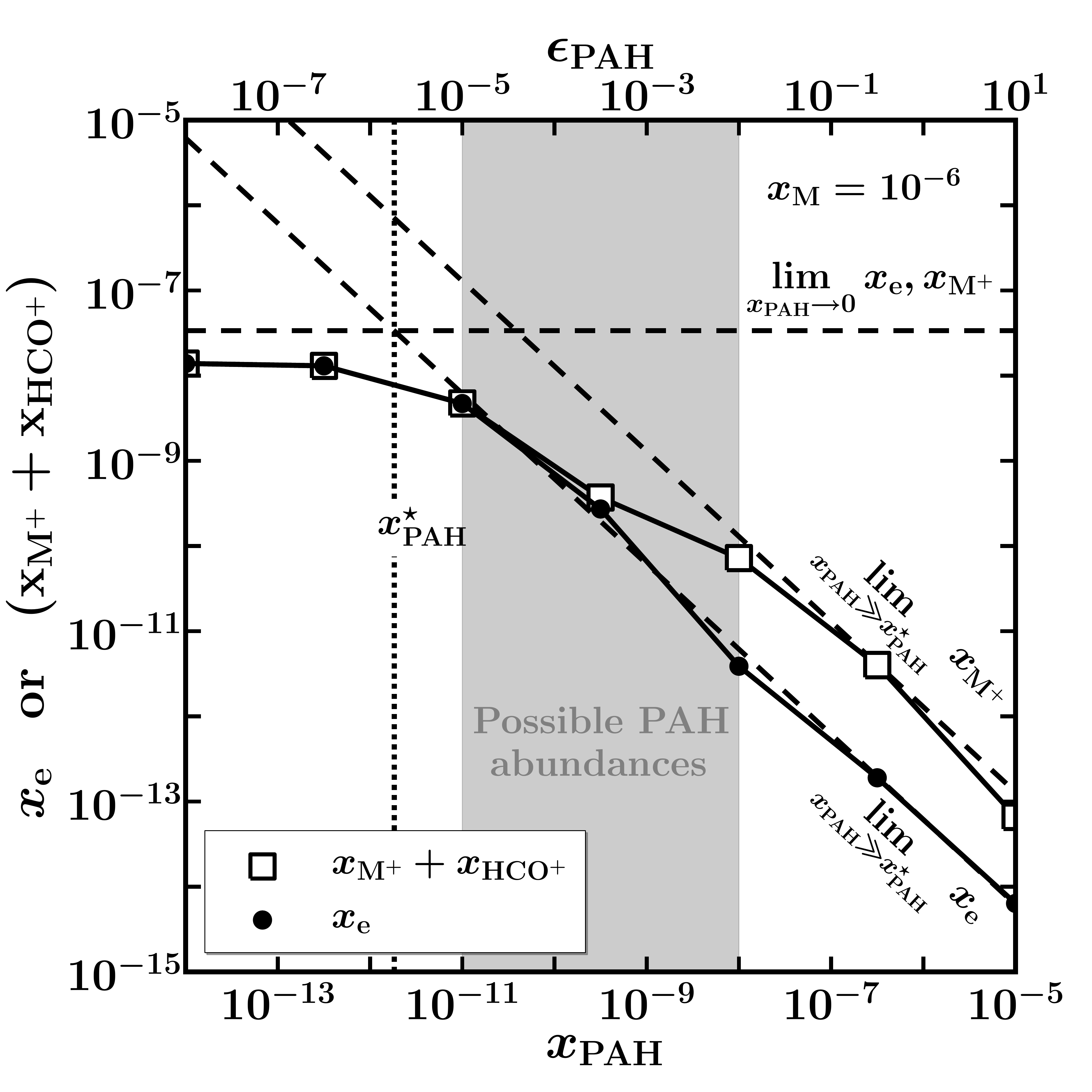}
\caption{Same as Figure \ref{fig_x_pah_crit_M8} but for the metal-rich
  case ($x_{\rm{M}}=10^{-6}$). The curves for $x_{\rm e}$ and $x_{\rm i}$ at high PAH abundance
  ($x_{\rm PAH} > x_{\rm PAH}^\star$, where $x_{\rm PAH}^\star$ is now
  given by Equation \ref{x_pah_crit_M6}) are essentially the same
  as in Figure \ref{fig_x_pah_crit_M8}: when PAHs dominate charge balance,
  the metal abundance ceases to matter.
 }
\label{fig_x_pah_crit_M6}
\end{figure}

\paragraph{Metal-rich case.}\label{mpc}
Analogous results are obtained for the metal-rich case as shown in Figure \ref{fig_x_pah_crit_M6}, with the only difference that M$^+$ replaces
${\rm HCO}^+$ as the dominant ion. In the limit of low PAH
abundance, the abundance of free charges is
\begin{eqnarray}
x_{\rm e} =x_{\rm{M^{+}}} & = & \sqrt{\zeta/ n_{\rm H_{2}}
  \alpha_{\rm{M^{+},e}}}  \label{low_pah_M6}\\ 
 & \sim & 3 \times 10^{-8} \left(\frac{L_{\rm{X}}}{10^{29} \,
    \rm{erg \,s^{-1}}}\right)^{1/2} \nonumber \\
& &\times \left( \frac{\Sigma}{0.3 \,{\rm g \, cm}^{-2}}
 \right)^{-1/2} \left( \frac{a}{3 \, {\rm AU}} \right)^{-0.5}\label{low_pah_M6_a}\\ \nonumber
\\
& &  {\rm \,\, for \,\, high \,\, metals \,\, and \,\, low \,\, PAHs}. \nonumber
\nonumber
\end{eqnarray}
In going from Equations (\ref{low_pah_M6}) to (\ref{low_pah_M6_a})
we account for the distance dependence of temperature but assume
material is optically thin to stellar X-rays.
Just as assuming all ions took the form of HCO$^{+}$ in the standard model
gave a lower limit (Equation \ref{low_pah}) for the ionization fraction, assuming
that all ions take the form of metals gives
an upper limit (Equation \ref{low_pah_M6}) because 
fast recombination of electrons with HCO$^+$ is neglected.

The analogous asymptotic solutions in the high PAH limit are practically
unchanged from Equations (\ref{el_high_pah}) and (\ref{ion_high_pah}) because
the charging rates of PAHs by HCO$^+$ and ${\rm M}^+$ are similar;
the mass of the HCO$^+$ molecule and that of a metal ion like Mg$^+$
are similar. The critical PAH abundance at which PAHs begin to reduce
the number of free charges is

\begin{eqnarray}
x_{\mathrm{PAH}}^{\star} & = & \sqrt{\frac{\zeta \alpha_{\rm
      M^{+},e}}{n_{\rm H_{2}} (\alpha_{\rm PAH,e})^2_{\langle Z
      \rangle = 0}}} \label{x_pah_crit_M6} \\
 & \sim & 2 \times 10^{-12} \left(\frac{L_{\rm{X}}}{10^{29} \,
    \rm{erg \,s^{-1}}}\right)^{1/2} \nonumber \\ \nonumber 
& & \times \left(\frac{\Sigma}{0.3 \,
    \rm{g \, cm^{-2}}}\right)^{-1/2} \left(\frac{a}{3
    \,\rm{AU}}\right)^{-0.2} \\ \nonumber
\\
& &  {\rm \,\, for \,\, high \,\, metals} \nonumber 
\end{eqnarray}
and is confirmed by the code.

\subsection{Degree of Magnetic Coupling: $Re$ and $Am$} 
\label{sec_AmRe}

We measure the extent of the MRI-active column by means of the
magnetic Reynolds number $Re$ (Equation \ref{Re}) and the ion-neutral
collisional frequency $Am$ (Equation \ref{Am}).  Figures \ref{fig_m8}
and \ref{fig_m6} show both dimensionless numbers as a function of the
surface density $\Sigma = N \mu$ penetrated by X-rays at $a = 3$ and
30 AU, respectively, over the range of observationally inferred PAH
abundances.  We overplot for comparison the solution obtained when we
omit PAHs completely. 

In both Figures \ref{fig_m8} and \ref{fig_m6},
the middle panels display results for our standard model
parameters: $x_{\rm M} = 10^{-8}$ per H$_2$, $L_{\rm X} = 10^{29}$ erg
s$^{-1}$, and $\zeta_{\rm CR} = 0$.  In each of the panels on the left
and on the right, we vary one of these parameters. We describe here
results for our standard model and compare with
other test cases in Section \ref{sec_lx}.

The middle top panels of Figures \ref{fig_m8} and \ref{fig_m6} show that if $Re$ were the only
discriminant, MRI-active surface layers could well exist, even with
PAHs present. At surface densities $\Sigma \sim 0.3$ g cm$^{-2}$
(column densities $N \sim 10^{23}$ cm$^{-2}$), $Re$ lies comfortably
above the critical values of $10^2$--$10^4$ (Section \ref{mri}) required
for plasma to couple to the magnetic field, for a wide range of
possible PAH abundances.  If the critical $Re \sim 10^2$ (as assumed
by BG), and if PAHs are at their lowest possible abundance as inferred
from observation ($\epsilon_{\rm PAH} = 10^{-5}$ relative to the ISM;
inverted triangles), then the MRI-active layer could extend as far as
$\Sigma \sim 20$ g cm$^{-2}$---if ohmic dissipation were the only
limiting factor for the MRI.

But ohmic dissipation is not the only factor. The same margin of
safety enjoyed by $Re$ does not at all apply to the ambipolar
diffusion number $Am$, for any surface density. Even in the
unrealistic case that there are no PAHs, $Am$ stays $<10$ in the
middle bottom panels of Figures \ref{fig_m8} and \ref{fig_m6}. By comparison,
values of $Am$ exceeding $10^2$ are reported by
\citet{Hawley:1998p5481} as necessary for the MRI to excite turbulence in
predominantly neutral gas. When PAHs are present, $Am$ barely exceeds
1, and then only for the low end of possible PAH abundances.  Compared
with ohmic dissipation, ambipolar diffusion seems the much greater
concern for the viability of the MRI in disk surface layers.

Values of $Am (\Sigma)$ and $Re(\Sigma)$ vary only slightly as the
stellocentric distance increases from $a = 3$ AU (Figure \ref{fig_m8},
middle) to 30 AU (Figure \ref{fig_m6}, middle); the former
decreases while the latter increases, each typically by factors of a
few. This behavior is readily understood. First recognize that $x_{\rm i}
\propto a^{-0.6}$ approximately; this is an average scaling between
the low PAH limit, which implies $x_{\rm i} \propto a^{-0.5}$
according to Equation (\ref{low_pah}), and the high PAH limit, which
implies $x_{\rm i} \propto a^{-5/7} \approx a^{-0.7}$ according to Equation
(\ref{ion_high_pah}).  Combining this result with $n_{\rm{H_2}}
\propto h^{-1} \propto \Omega/T^{1/2} \propto a^{-9/7}$, we find that
$Am = x_{\rm i}n_{\rm H_2}/\Omega\propto a^{-0.4}$.  Similarly, $Re =
c_{\rm s} h / D \propto x_{\rm e}T^{1/2}/\Omega \propto a^{0.7}$.

In computing $Am$, we have omitted the contribution from collisions
between neutral H$_2$ and negatively charged PAHs. The latter are as
well coupled to magnetic fields as molecular ions are---see, e.g., the
ion and grain Hall parameters calculated in Section 2.2 of BG. Thus,
collisions between H$_2$ and charged PAHs should increase $Am$. 
However, we find that in practice the gain is negligible. We
estimate that the collisional rate coefficient $\beta_{\rm in}$ that
enters into $Am$ is about the same for charged PAHs as for ions; in
both cases a collision with an H$_2$ molecule is mediated by the
induced dipole in H$_2$, and the relative velocity is dominated by the
thermal speed of H$_2$.  At $\Sigma \gtrsim 10$ g cm$^{-2}$, charged
PAHs are about as abundant as ions and thus raise $Am$ by a factor of
2---but at these $\Sigma$'s, $Am$ is already too low for the MRI to be
viable.  At $\Sigma \lesssim 10$ g cm$^{-2}$---i.e., at those columns
where $Am$ peaks---charged PAHs are much less abundant than ions and
thus hardly affect $Am$.

\subsubsection{Higher $L_{\rm{X}}$ and $T_{\rm X}$, Higher and Lower $x_{\rm M}$, and Cosmic-ray Ionization} 
\label{sec_lx}

\begin{figure*}[h!]
\epsscale{1.2}
\plotone{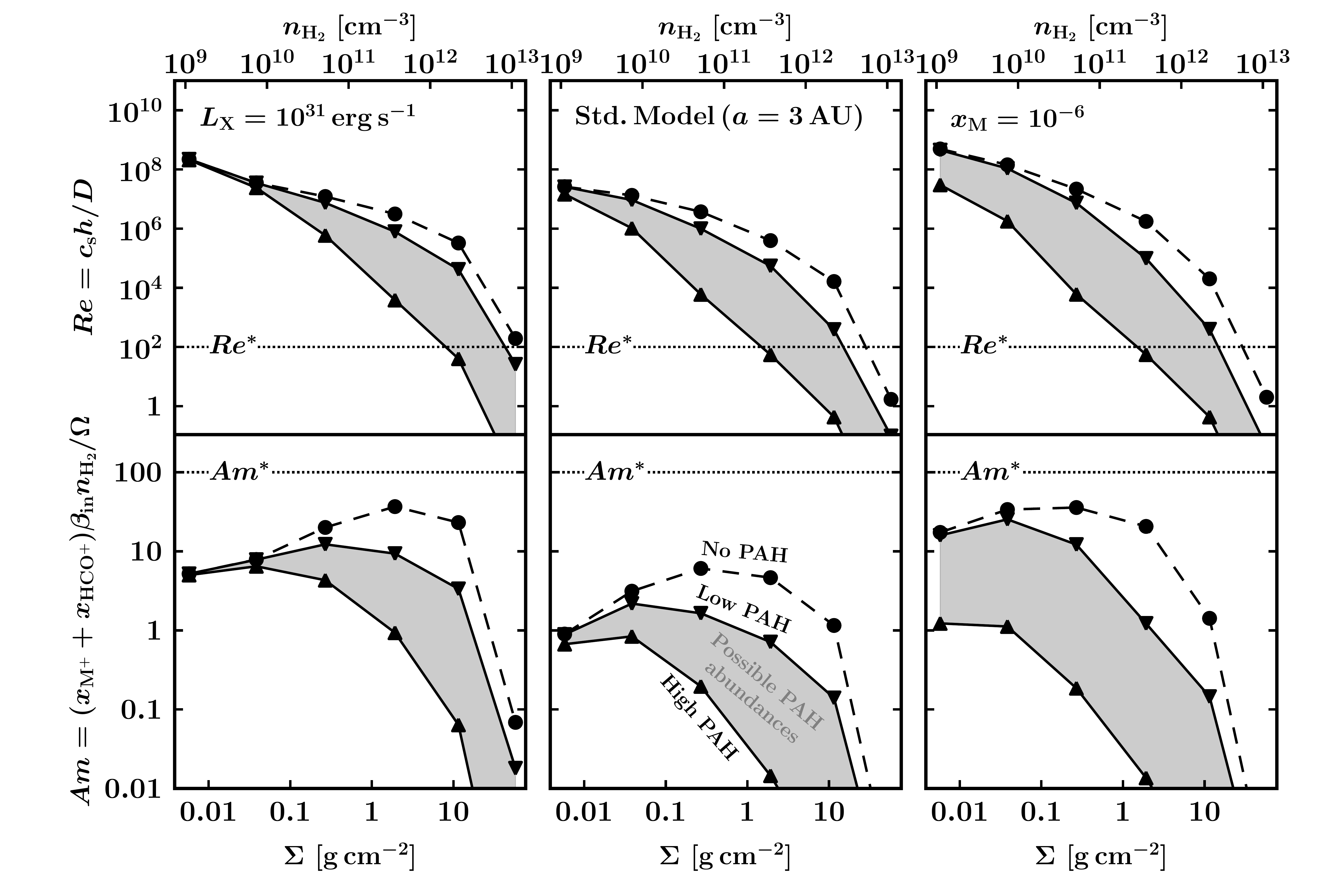}
\caption{Magnetic Reynolds number $Re$ and ambipolar diffusion number
  $Am$ as a function of surface density $\Sigma$ at $a=3$ AU. The
  middle panels show results for our standard model
  ($x_{\rm{M}}=10^{-8}$, $L_{\rm{X}} = 10^{29}$ erg s$^{-1}$,
  $\zeta_{\rm{CR}}=0$). The side panels have the same parameters as our
  standard model, except for a 100$\times$ more luminous X-ray source
  (left panels), and a 100$\times$ greater metal abundance (right
  panels).  Oppositely pointing triangles bracket values for $Am$ and
  $Re$ corresponding to possible PAH abundances. These abundances were
  inferred in Section \ref{sec_PAHs} from observations. The dashed curve
  refers to the case with no PAHs and is shown for comparison
  only. Polycyclic aromatic hydrocarbons reduce ionization fractions
  and thus the degree of magnetic coupling by an order of magnitude or
  more.
  The dotted lines mark the critical values $Am^*$ and $Re^*$ above which
  coupling between magnetic fields and neutral gas is sufficient to
  drive the MRI (Section \ref{mri}).  The curves for $Am$ first rise as
  $\Sigma$ increases---a consequence of the increasing number
  density---and then fall as the ion fraction decreases, never
  reaching $Am^*$.  Ambipolar diffusion threatens the MRI more than ohmic dissipation does. If the critical $Am$ required for good
  collisional coupling between ions and neutrals is $Am^*=10^2$, as
  evidenced in simulations by \citet{Hawley:1998p5481}, then even a
  disk without PAHs cannot sustain X-ray-driven MRI at any $\Sigma$.}
\label{fig_m8}
\end{figure*}

\begin{figure*}[h!]
\epsscale{1.2}
\plotone{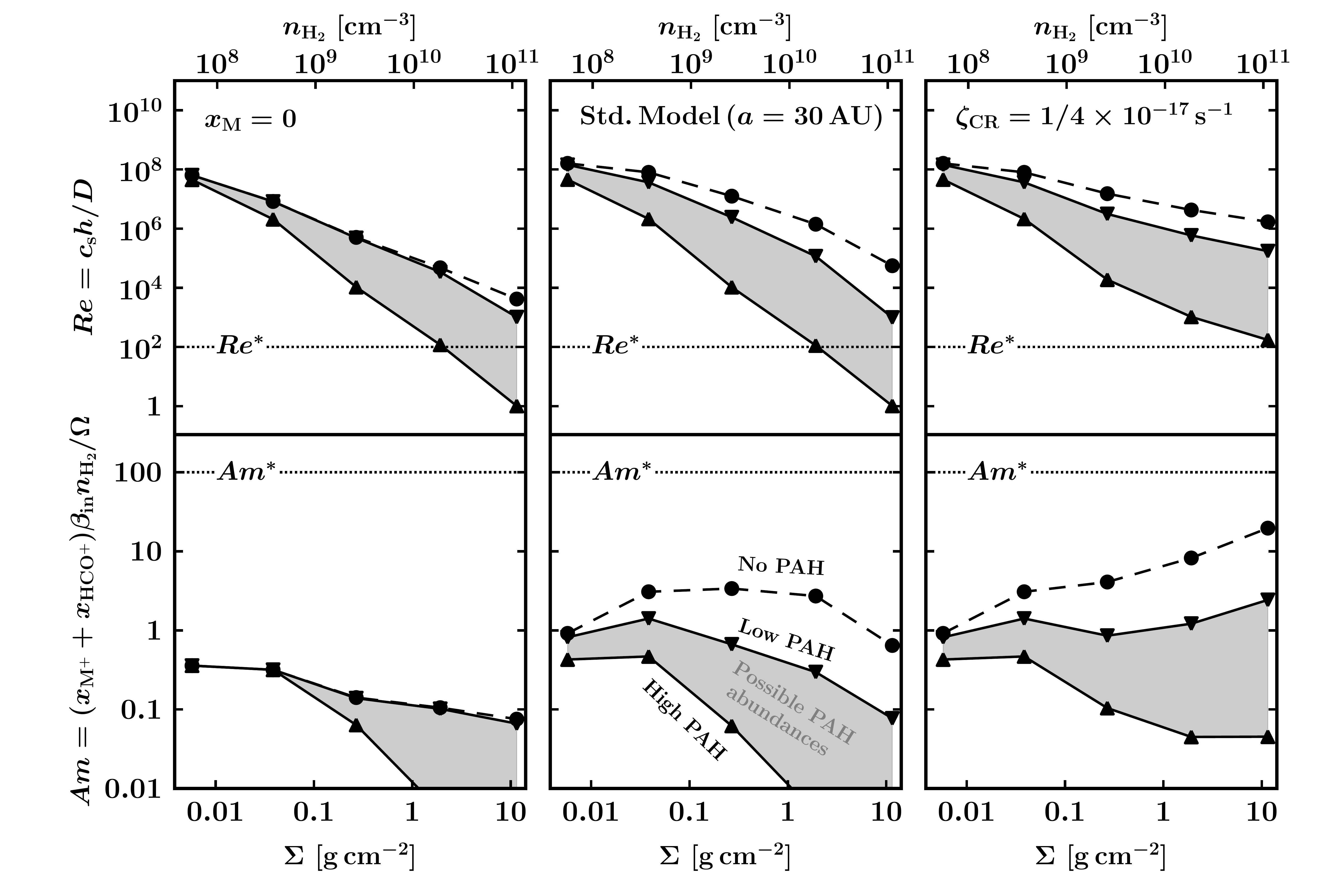}
\caption{Same as Figure \ref{fig_m8}, but at a stellocentric distance
  of $a=30$ AU. The middle panels show results for our standard model
  ($x_{\rm{M}}=10^{-8}$, $L_{\rm{X}} = 10^{29}$ erg s$^{-1}$,
  $\zeta_{\rm{CR}}=0$). The side panels have the same parameters as our
  standard model, except that gas-phase metals are omitted in the left
  panels ($x_{\rm{M}}=0$), and sideways cosmic-rays are added in the
  right panels ($\zeta_{\rm{CR}}=1/4 \times 10^{-17}$ s$^{-1}$).}
\label{fig_m6}
\end{figure*}

In each of the leftmost and rightmost panels of Figures \ref{fig_m8} and \ref{fig_m6},
we vary one model parameter away from its standard value. We begin
with the case of higher $L_{\rm X}$. Increasing $L_{\rm{X}}$ certainly raises $Re$ and $Am$, but as the
left panels of Figure \ref{fig_m8} show, even a fairly high $L_{\rm{X}} = 10^{31}$ erg s$^{-1}$
only causes $Am$ to just exceed 10 at the lowest PAH abundance. In the
limit of low PAH abundance, $x_{\rm e} = x_{\rm i} \propto L_{\rm{X}}^{1/2}$,
as predicted by Equation (\ref{low_pah}). In the limit of high PAH
abundance, $x_{\rm e}$ and $x_{\rm i}$ scale linearly with $L_{\rm{X}}$,
according to Equations (\ref{el_high_pah}) and (\ref{ion_high_pah}).
Thus the space of possible values of $Re$ and $Am$ narrows with increasing
$L_{\rm{X}}$, as the lower envelope increases as $L_{\rm{X}}$ while the upper
envelope increases as $L_{\rm{X}}^{1/2}$.

These same scaling relations, with $L_{\rm X}$ replaced by the
ionization rate $\zeta$ at fixed distance, enable us to estimate the
effects of a higher $T_{\rm X}$, a case we did not explicitly compute
using our numerical model.  According to IG, raising $kT_{\rm X}$ from
3 keV (our standard value) to 8 keV increases the ionization rate
$\zeta$ by factors of 2--4 at $\Sigma = 1$--30 g cm$^{-2}$. Thus in
the extreme case that $L_{\rm X} = 10^{31}$ erg s$^{-1}$ and $kT_{\rm
  X} = 8$ keV---parameters appropriate only for a small minority of
young stars (\citealt{Telleschi:2007p7899};
\citealt{Preibisch:2005p6831})---we apply the low-PAH scaling
 relation $x_{\rm i} \propto \zeta^{1/2}$ to the
bottom left panel of Figure \ref{fig_m8} to find that the largest
possible value of $Am$ is $\sim$20, obtained if PAHs are at their
lowest plausible abundance.

The rightmost panels of Figure \ref{fig_m8} display the case
of a higher metal abundance $x_{\rm M} = 10^{-6}$ per H$_2$.
As discussed in Section \ref{sec_metals}, the higher metal abundance is not especially
realistic and is considered primarily as an exercise. Comparing
the middle and right panels of Figure \ref{fig_m8}, we see that increasing the
metal abundance by a factor of 100 raises $Re$ and $Am$ at low PAH
abundance by a factor of $\sim$10.  At high PAH abundance, $Re$ and
$Am$ also increase with increasing metal abundance, but the gain is
less. This same behavior is reflected in the solid curves of Figures
\ref{fig_x_pah_crit_M8} and \ref{fig_x_pah_crit_M6}: at low PAH
abundance, increasing $x_{\rm M}$ by a factor of 100 leads to a factor
of $\sim$10 increase in $x_{\rm e} = x_{\rm i}$, but at high PAH
abundance, the now divergent curves for $x_{\rm e}$ and $x_{\rm i}$
are essentially independent of metal abundance. Equations
(\ref{el_high_pah}) and (\ref{ion_high_pah}) from our analytic
analysis reflect this insensitivity to metal abundance at high PAH
abundance.

At $\Sigma \gtrsim 1$ g cm$^{-2}$, gas temperatures may be so low that
all of the metals condense onto grains. The case $x_{\rm M} = 0$ is
shown in the leftmost panels of Figure \ref{fig_m6}.  Here $Am
\lesssim 0.1$ for $\Sigma \gtrsim 1$ g cm$^{-2}$, and it seems
safe to conclude that X-ray driven MRI is unviable under these conditions.
Additional ionization by ``sideways cosmic-rays'' at $a=30$ AU is
considered in the rightmost panels of Figure \ref{fig_m6}. These
cosmic-rays, which we have imagined enter the disk edge-on from the
outside, dominate stellar X-rays at large $\Sigma$.  At $\Sigma \sim
10$ g cm$^{-2}$---comparable to the full surface density of the disk
at $a = 30$ AU---sideways cosmic-rays raise the maximum value of $Am$
to $\sim$2.  We have verified that the gains in $Am$ afforded by
cosmic-rays are consistent with our scalings of $x_{\rm e}$ and
$x_{\rm i}$ with $\zeta$ as derived above.

\subsubsection{Chemical Equilibration Timescales vs. Dynamical Timescales} 
\label{sec_timescale}

In assessing whether disk surface layers are MRI-active, we have
relied on the critical value $Am^\ast \sim 10^2$ reported by HS. As a
simplifying assumption, HS held fixed the global (box-integrated) ion
abundance in each of their simulations. A fixed ion abundance would
apply if the chemical equilibration timescale $t_{\rm eq}$ exceeds the
dynamical timescale $t_{\rm dyn} = \Omega^{-1}$ over which HS's
simulations ran. A fixed ion abundance would also apply if ion
recombination occurs predominantly on condensates, regardless of
$t_{\rm eq}/t_{\rm dyn}$ (e.g., \citealt{MacLow:1995p7143}).  This last
statement follows from our Equation (\ref{ion_high_pah}), which shows $x_{\rm
  i}n_{\rm H_2}$ does not depend on $n_{\rm H_2}$ in the high
condensate limit.

The high condensate limit applies for PAH abundances near the high end
of those inferred from observation (Figure
\ref{fig_x_pah_crit_M8}). For this high PAH case we expect the
assumption of constant ion abundance, and by extension the results of
HS, to hold.  For high PAH abundance and our standard X-ray
luminosity, $Am < 1$ for all $\Sigma$ and $a$ (Figure \ref{fig_m8}),
and our conclusion that X-ray-driven MRI shuts down everywhere seems safe.

\begin{figure}[h!]
\epsscale{1.0}
\plotone{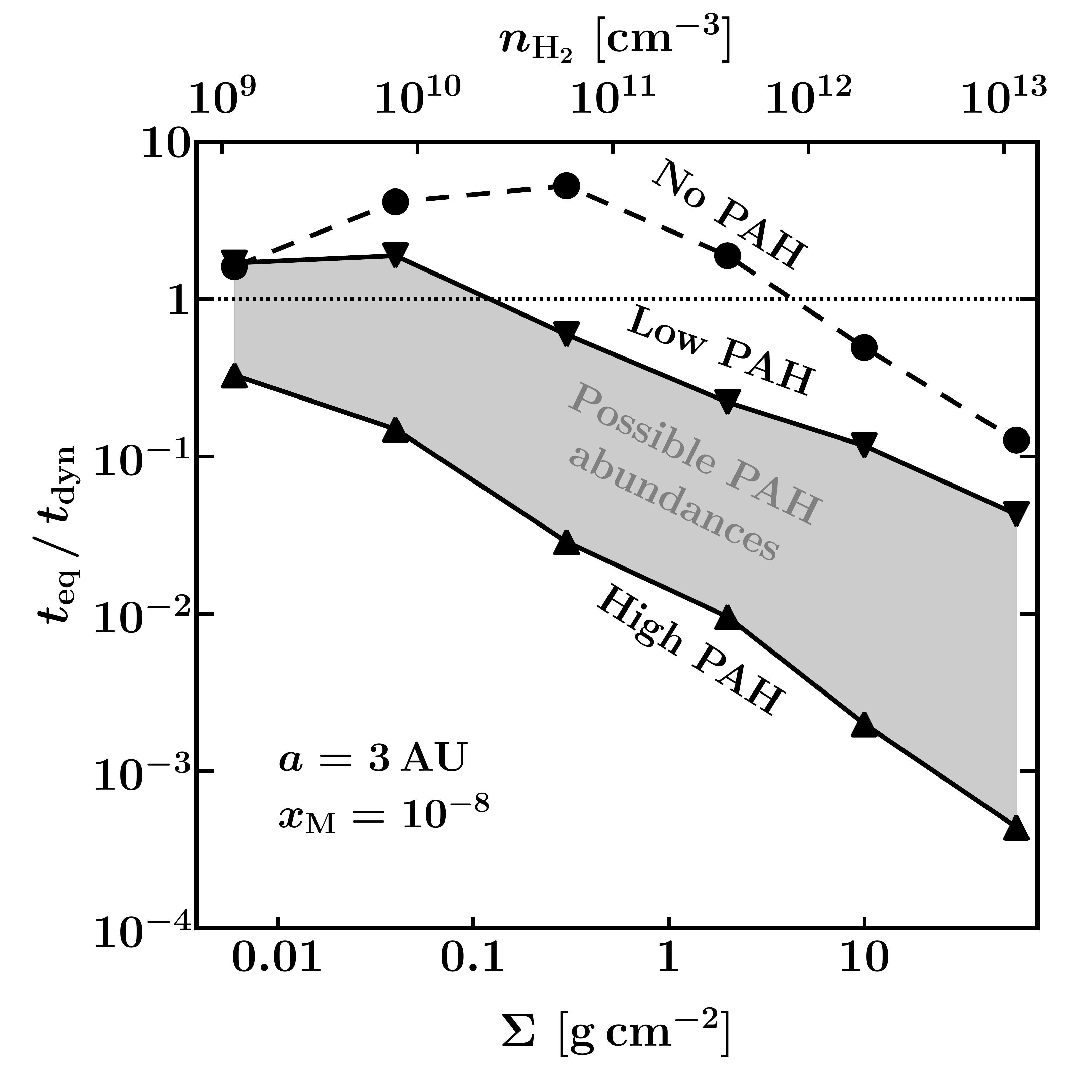}
\caption{Ratio of the chemical equilibration timescale $t_{\rm{eq}}$,
  computed according to the procedure described in
  Section \ref{sec_numerical}, to the dynamical time $t_{\rm{dyn}} =
  \Omega^{-1}$, for $a = 3$ AU and $x_{\rm{M}}=10^{-8}$.  Values of
  $t_{\rm eq}/t_{\rm dyn}$ at $a=30$ AU are typically lower than those
  shown here by factors of 3 or less. Simulations by HS assumed a
  constant (volume-integrated) abundance of ions, a condition
  satisfied if $t_{\rm{eq}}/t_{\rm{dyn}} > 1$. The ion abundance is
  also constant if ion recombination occurs primarily on condensates
  (see Equation \ref{ion_high_pah}), a situation that obtains for PAH
  abundances near the high end of those inferred from observations.
  For the lowest possible PAH abundances, $0.1 <
  t_{\rm{eq}}/t_{\rm{dyn}} < 1$ at $\Sigma = 0.1$--10 g cm$^{-2}$. In
  this low PAH case, the results of HS might still be expected to
  apply to order unity. Even if they do not, we argue in the main text
  that when $t_{\rm eq}/t_{\rm dyn} < 1$ the effects of ambipolar
  diffusion should be even stronger than reported by HS.}
\label{fig_timescales}
\end{figure}

For PAH abundances at the low end of those inferred from observation,
Equation (\ref{ion_high_pah}) for the high condensate limit does not
apply. Moreover, as shown in Figure \ref{fig_timescales}, $t_{\rm
  eq}/t_{\rm dyn} < 1$---but only by a factor of 10 at most for the
low PAH case and for $\Sigma \lesssim 10$ g cm$^{-2}$.  Because
$t_{\rm eq}/t_{\rm dyn} \gtrsim 0.1$ under these conditions, the
assumption of constant ion abundance, although not strictly valid,
might still be good enough that the results of HS hold to order
unity. But even if they do not, we would argue that the sign of any
correction for a dynamically variable ion abundance would only hurt
the prospects for MRI turbulence.
In the simulations of HS---see also
\citet{Brandenburg:1994p7177} and \citet{MacLow:1995p7143}---ions
became concentrated in thin filaments within magnetic
nulls. Recombination rates inside the dense filaments were higher than
those outside.  Were such simulations to account for ion
recombination, lower ion densities within the filaments would result,
and neutrals would be even less coupled to ions.
Consequently, $Am^\ast$ would be even higher than the reported
value of $\sim$$10^2$.

 \vspace{0.3in}
\subsection{Comparison with Previous Work: Ionization Fractions}\label{compare}

The ionization chemistry in disks remains inherently uncertain, with
rate coefficients for many reactions in the UMIST
database determined to no better than factors of $\sim$3 \citep{Vasyunin:2008p6845}. A measure of the uncertainty in the ionization fraction
in disks is given by the differences between the simple and complex networks
computed by BG, which amount to factors of 2--10 for the electron
fraction $x_{\rm e}$.  We should
reproduce their results by at least
this margin, as a validation of our code. In the following we directly
compare our results to those of BG and TCS, adjusting the input
parameters of our code to match theirs.
Once we match these input parameters, any difference in
our codes' outputs should result primarily from our different
chemical networks (ours is the simplest of the three),
and not from differences in radiative transfer, as
all our codes rely on the ionization rates calculated
by \citet{Igea:1999p2931}.

\begin{figure}[h!]
\epsscale{1.0}
\plotone{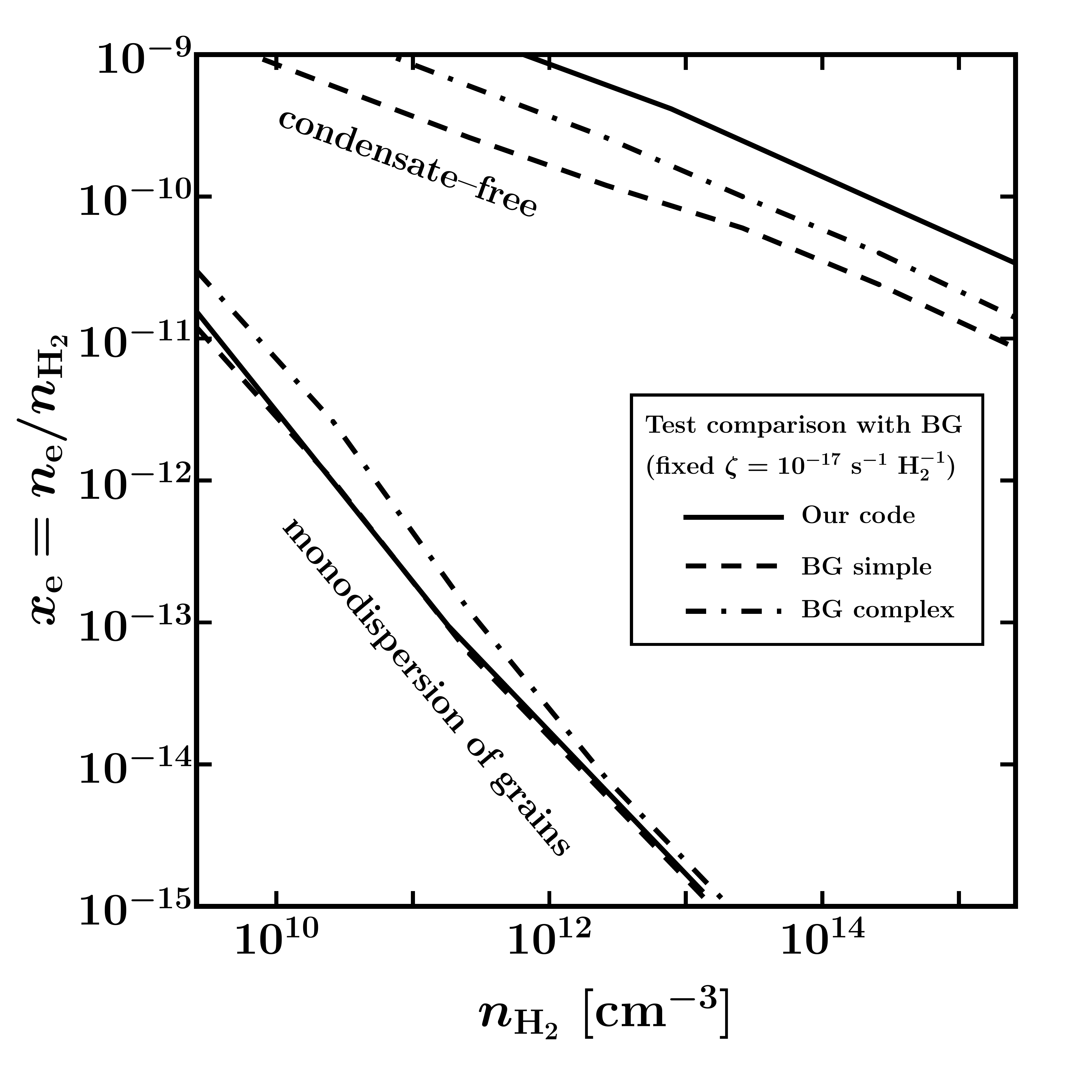}
\caption{Test comparison with BG: electron abundance as a function of
  gas density at a fixed ionization rate of $\zeta=10^{-17}$ s$^{-1}$
  H$_2^{-1}$. The upper set of lines are for a condensate-free system,
  and the lower set are for a monodispersion of grains. Results from
  BG were drawn from their Figure 3. To generate our results, the
  parameters of our code were reset to those of BG: temperature
  $T=280$ K, metal abundance $x_{\rm{M}} = 2.5 \times 10^{-8}$ per
  H$_2$, electron-grain sticking coefficient $S_{\rm{e}}=0.03$, grain
  radius $s=0.1 \, \mu$m, internal grain density $\rho_s = 3$ g
  cm$^{-3}$, and a mass fraction in grains relative to gas of 1\%.}
\label{fig_BG}
\end{figure}

We start with BG by computing $x_{\rm e}$ as a function of density
$n_{\rm H_2}$ at a fixed ionization rate $\zeta = 10^{-17}$ s$^{-1}$,
following their Figure 3.  We reset $T = 280$ K, $x_{\rm M} = 2.5
\times 10^{-8}$ per H$_2$, and the electron-grain sticking coefficient
$S_{\rm e} = 0.03$ to match their standard parameters.  To compare to
their ``grain-free'' case, we run our code without any PAHs or
grains. To compare to their standard monodispersion of grains, we run
our code with a single population of grains having $s = 0.1 \,\mu$m,
internal density $\rho_s = 3$ g cm$^{-3}$, and a mass fraction of 1\%
relative to gas.  Figure \ref{fig_BG} shows the comparison. Our
results for the condensate-free case track those of BG, but are higher
by factors of 3--10 depending on whether the comparison is made with
their simple or complex network.  For the case with grains, the
agreement with the simple model is excellent and that with the complex
model is good to a factor of 2.

\begin{figure}[h!]
\epsscale{1.0}
\plotone{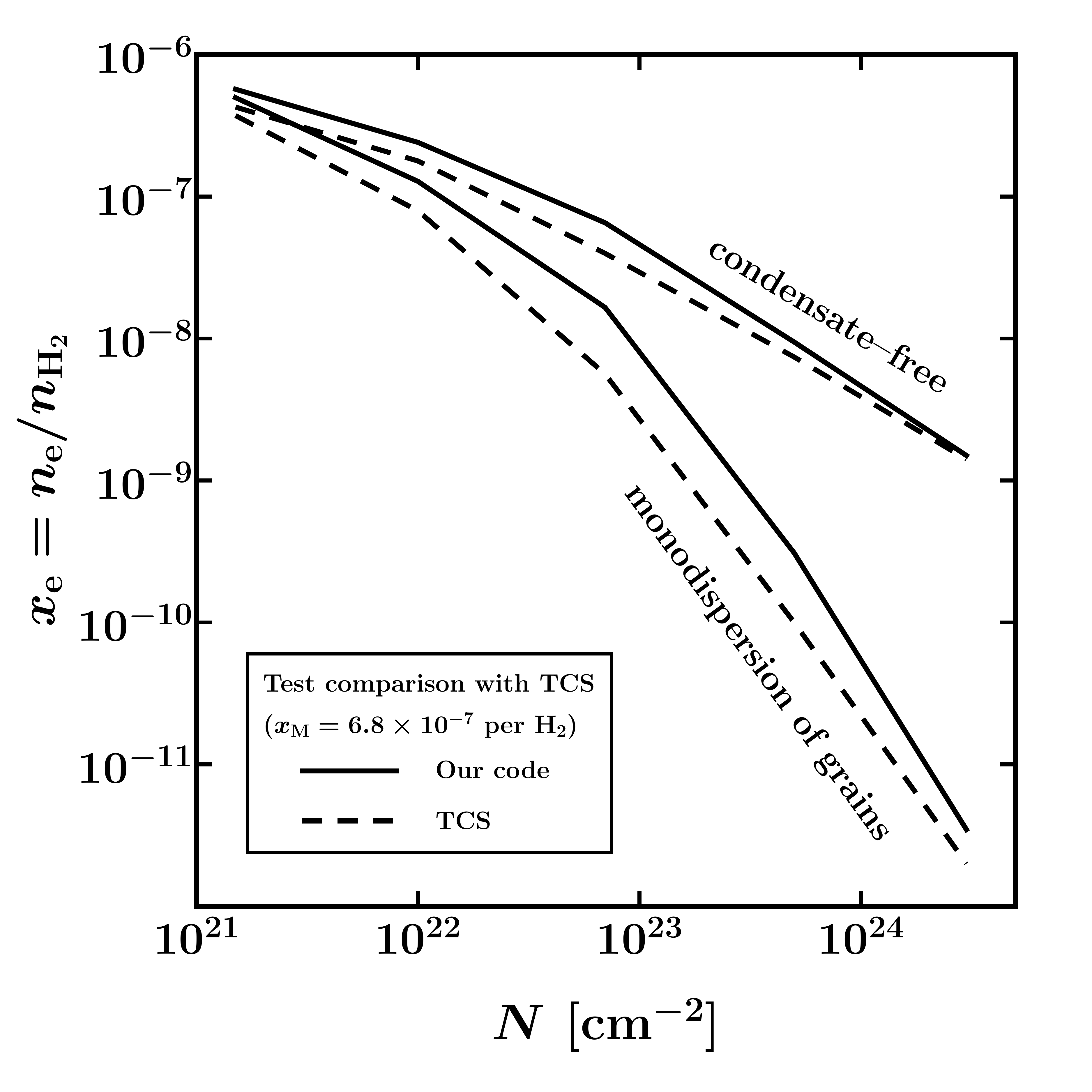}
\caption{Test comparison with TCS: electron abundance as a function of
  column depth penetrated by X-rays. The upper pair of lines are for a
  condensate-free system, and the lower pair are for a monodispersion
  of grains. Ionization rates are from IG. Results from TCS are taken
  from their Figure 1. To generate our results, we reset the
  parameters of our code to match those of TCS: stellocentric distance
  $a=5$ AU, X-ray luminosity $L_{\rm{X}}= 2 \times 10^{30}$ erg
  s$^{-1}$, temperature $T=125$ K, metal (magnesium) abundance
  $x_{\rm{M}} = 6.8 \times 10^{-7}$ per H$_2$, electron-grain sticking
  coefficient $S_{\rm{e}}=0.03$, grain radius $s=1 \, \mu$m, internal
  grain density $\rho_s = 5$ g cm$^{-3}$, and a mass fraction in
  grains relative to gas of 1\%.}
\label{fig_TCS}
\end{figure}

In Figure \ref{fig_TCS}, we make a similar comparison with TCS,
computing $x_{\rm e}$ as a function of $N$ at a distance of $a = 5$ AU
from an X-ray source of $L_{\rm{X}} = 2 \times 10^{30}$ erg s$^{-1}$ and
$kT_{X} = 5$ keV, for $T = 125$ K and a metal abundance of $x_{\rm M}
= 6.8 \times 10^{-7}$ per H$_2$. We consider the two cases of their
Figure 1, one without any grains or PAHs, and another with a single
population of grains having $s = 1 \, \mu$m, $\rho_s = 5$ g cm$^{-3}$,
and a mass fraction of 1\%. For both cases our computed
electron abundances are higher, but only by factors
of 2 or less.

These comparisons with BG and TCS give us confidence that we have
computed ionization fractions about as well as they did. Where our
ionization fractions differ, ours are often higher.
Our higher values will only bolster the conclusion we make in
Section \ref{sec_discussion} that thicknesses of X-ray-ionized
MRI-active surface layers
have been overestimated by them and others.

\section{SUMMARY AND DISCUSSION}
\label{sec_discussion}

In Section \ref{intro}, we presented the evidence that holes and gaps of
transitional disks are cleared by companions to their host stars.
Residing within the hole, these companions 
could either be stars---already observed in about
half of all transitional systems---or multi-planet systems.
A single Jupiter-mass planet on a circular orbit
carves out too narrow a gap to explain the large cavities inferred
from observations. But multiple planets can
shuttle gas quickly from one planet to the next, all the way down to
the central star. Surface densities fall in inverse proportion to
radial infall speeds, and radial infall speeds can approach freefall
speeds for sufficiently many and massive planets.
In this way, multi-planet systems might help to clear
holes and simultaneously sustain stellar accretion rates that approach
those in disks without holes.  The more eccentric the planets' orbits,
the fewer of them may be required to explain a given hole
size. Accretion in the presence of multi-planet systems has not
received much attention and seems an interesting area for future
simulation (e.g., Zhu et al.~2010, submitted).

Stellar or planetary companions regulate accretion velocities $v$ but
do not give rise to mass accretion rates $\dot{M}$ in the first place.
A planet orbiting just inside the circumference of a disk hole exerts
torques to repel gas in the hole's rim away from the star.  Thus, the
shepherding planet may reduce $\dot{M}$---and indeed accretion rates
in transitional systems tend to be smaller than those in conventional
disks \citep{Najita:2007p6175}---but the planet does not initiate disk
accretion. A separate mechanism must act to
pull or diffuse gas inward from the hole rim to supply the stellar
accretion rates that are observed.  That mechanism may be turbulence
driven by the magnetorotational instability (MRI), activated by
stellar radiation ionizing rim gas.  Whether the MRI can operate depends
on how well ionized the gas is. The greater the free electron
fraction, the greater the magnetic Reynolds number $Re$, and the less
ohmic dissipation dampens the MRI. The greater the atomic and
molecular ion densities, the greater the collisional rate
$Am$ between neutral particles and ions, and the less
ambipolar diffusion weakens the MRI.

A principal threat to the MRI is posed by dust grains, which
adsorb electrons and ions.  The smallest grains may present the
biggest danger, because in many particle size distributions the
smallest grains have the greatest surface area for
attachment.
The smallest grains that can also be detected observationally
are polycyclic aromatic hydrocarbons (PAHs), each several angstroms
across and containing of order a hundred carbon atoms. These
macromolecules may reside in the very disk surface layers that promise to
be MRI-active.  Excited by soft ultraviolet radiation, PAHs fluoresce
in a distinctive set of infrared emission lines detectable from {\it
  Spitzer} and from the ground.  The hydrocarbon molecules are
probably generated locally, photo-sputtered off larger particles
exposed to hard UV and X-ray radiation from host stars.

To assess the impact of PAHs on the MRI, we need to know PAH
abundances relative to gas. These can be inferred from observed PAH
emission lines.  Unfortunately such inferences are model dependent;
they depend on knowing the local grain opacity, because the soft UV
radiation that excites PAHs is also absorbed by grains. In other
words, observed PAH line intensities 
depend on PAH-to-dust ratios. It follows that the quantity of interest
to us---the PAH-to-gas ratio---depends on knowing the dust-to-gas
ratio. The latter can vary widely with the degree to which grains
settle toward disk midplanes.  The more grains have settled, the lower
are local dust-to-gas ratios, and the lower the PAH-to-gas abundance
that is needed to explain a given set of PAH emission spectra.  By
compiling a few lines of model-dependent evidence from the
literature, we estimated that disk PAHs have abundances anywhere from
$10^{-11}$--$10^{-8}$ per H$_2$, with lower values corresponding to a
greater degree of dust settling.

Such PAH abundances, although depleted relative to the ISM by
$10^{-5}$--$10^{-2}$, are still large enough to significantly weaken
the MRI in disk surface layers. In fact, they might even shut off X-ray-driven
MRI altogether, everywhere.  For stellar X-ray luminosities of
$L_{\rm{X}} = 10^{29}$--10$^{31}$ erg s$^{-1}$ and X-ray stopping
columns of $\Sigma \sim 1$--10 g cm$^{-2}$, PAHs reduce electron and
ion densities---which are not equal when PAHs are present---by factors
of $\sim$10 or more.  At these surface densities, the collisional
coupling frequency $Am \approx 10^{-3}$--10, depending on PAH
abundance and X-ray luminosity.  These values fall short, by 1--5 orders
of magnitude, of the critical value $Am^\ast \sim 10^2$
required for good coupling between ions and neutrals, as measured in
simulations by \citet[][HS]{Hawley:1998p5481}.
The potentially catastrophic effect that small grains can have on the
MRI was highlighted by Bai \& Goodman (2009, BG). Our study grounds
their concern in real-life observations.

Other studies reported X-ray-driven MRI-active surface layers to be
alive and well (e.g., Chiang \& Murray-Clay 2007, CMC; Turner,
Carballido, \& Sano 2010, TCS; and BG, in many of whose models the
active layer extended to $\sim$1 g cm$^{-2}$, even with grains
present). We should understand why our conclusions differ from
theirs. In part, the difference arises because previous studies
neglected PAHs. A further difference with TCS is that they assumed a
metal abundance of $x_{\rm M} = 6.8 \times 10^{-7}$ per H$_2$, nearly
2 orders of magnitude higher than our standard model value, and one
that we find difficult to justify.  Still another difference, as
significant as any of the ones just mentioned, is the criterion used
for whether ambipolar diffusion defeats the
MRI. \citet{Turner:2010p6693} assumed $Am^\ast \sim 1$ (see their
Equation 9).  \citet{\BG} did not present results for $Am$. Using
their data, we computed the $Am$ values characterizing their claimed
active layers. At $a \gtrsim 1$ AU, the $Am$ values of BG's grain-free
active layer are at most on the order of unity. For BG's standard
models containing grains, $Am \approx 0.0004$--0.4, with the lowest
value corresponding to $a= 50$ AU and a population of grains having
two sizes, and the highest value corresponding to $a= 1$ AU and a
single-sized grain population (we computed both limits using results
from their complex chemical network). \citet{Hawley:1998p5481} showed
that when $Am \lesssim 0.01$, ions and neutrals were effectively
decoupled. Even when $Am \sim 1$, HS showed that the MRI saturation
amplitude scaled with the ion and not the neutral density, with the
neutrals acting to damp out MRI turbulence in the ions. If the MRI
drives turbulence only in the ions of protoplanetary disks, it might
as well not operate at all, given how overwhelmingly neutral such
disks are.

\subsection{Future Directions}\label{future}
We have shown in this paper that the MRI cannot drive surface
layer accretion under typical circumstances in protoplanetary disks, either
transitional or conventional, if the critical $Am^\ast \sim 10^2$
and if stellar X-rays and Galactic cosmic-rays are the dominant
source of ionization. These two ``if''s are subject
to further investigation. We discuss each in turn.

We are not aware of more modern estimates of $Am^\ast$ apart from that
given by HS. As these authors cautioned, numerical resolution is a
greater concern for two-fluid simulations than for single-fluid ones,
and HS did not demonstrate convergence of their results with
resolution. In addition, the value of $Am^\ast$ was not as precisely
determined by HS for toroidal field geometries as for vertical
ones---although $Am^{\ast} \sim 10^2$ did seem to apply equally well
to the cases of uniform vertical field and zero net vertical field. 
Perhaps higher resolution simulations will reveal that
$Am^\ast < 10^2$---although accounting for ion recombination in these
simulations should only increase $Am^\ast$ (Section \ref{sec_timescale}).

The second possibility is that our model has neglected a
significant source of ionization. Stellar radiation just longward
of the Lyman limit---so-called far ultraviolet (FUV) radiation at
photon energies between $\sim$6 and 13.6 eV---can ionize trace species
such as C, S, Mg, Si, and Al (e.g., \citealt{Cruddace:1974p7788}). Of these
elements, C and S may be the most important, as they are the most
abundant and possibly least depleted onto grains. For example, if the
full solar abundance of C were singly ionized, the ion fraction
$x_{\rm i}$ would be a few $\times$ $10^{-4}$, or 5 orders of
magnitude higher than the largest values of $x_{\rm i}$ reported in this
paper!  At disk midplanes which are shielded from photodissociating
radiation, an order-unity fraction of the full solar abundance of C is
expected to take the form of CO ($x_{\rm CO} = 10^{-4}$;
\citealt{Aikawa:1996p3361}).  As computed in chemical models by 
\citeauthor*{Gorti:2004p7706} (2004; see also
\citealt{Tielens:1985p7751} and \citealt{Kaufman:1999p7775}),
CO near disk surfaces photodissociates nearly entirely
by FUV radiation into a layer of neutral C. At the highest altitudes,
nearly all of this carbon is photoionized by FUV radiation. The column density of C$^+$
depends on how many small grains having sizes $\lesssim 0.1 \mu$m are
present, as grains compete to absorb the same FUV photons that
photodissociate CO and photoionize C.

We may estimate maximum FUV-ionized column densities by
neglecting such dust extinction, and by neglecting shielding of FUV
radiation by molecular hydrogen. Consider a trace species T whose total number
density regardless of ionization state is $f_{\rm T}n_{\rm H_2}$. Take
all of T to be singly ionized within a Str\"omgren slab at the disk
surface: $n_{\rm T^+} = n_{\rm e} = f_{\rm T} n_{\rm H_2}$.
Per unit surface area of slab, the rate of photoionizations balances the rate
of radiative recombinations:
\begin{eqnarray}
\frac{L_{\rm FUV}}{E_{\rm FUV} 4 \pi a^2} & \sim & n_{\rm T^+}n_{\rm e} \alpha_{\rm T^+,e} h \nonumber \\
 & \sim & f_{\rm T}^2 n_{\rm H_2}^2 \alpha_{\rm T^+,e} h \,,
\end{eqnarray}
where the FUV luminosity capable of ionizing T is $L_{\rm FUV} \sim
10^{30}$ erg s$^{-1}$ \citep{Gorti:2009p7687}, the
photon energy $E_{\rm FUV} \sim 10$ eV, the rate coefficient
$\alpha_{\rm T^+,e} \sim 4 \times 10^{-12}$ cm$^3$ s$^{-1}$ at an
FUV-heated gas temperature of 300 K, and the slab thickness $h \sim 0.1
a$. Solve for the hydrogen column
\begin{eqnarray}
N_{\rm FUV} & = & n_{\rm H_2} h  \\
 & \sim & 2 \times 10^{22} \left( \frac{L_{\rm FUV}}{10^{30} \, {\rm
       erg}\, {\rm s}^{-1}} \right)^{1/2} \nonumber \\
& & \times \left( \frac{3 \,{\rm AU}}{a} \right)^{1/2}\left(
  \frac{10^{-4}}{f_{\rm T}} \right) \, {\rm cm}^{-2} \,, \nonumber
\end{eqnarray}
or equivalently
\begin{eqnarray} \label{eq_fuv}
\Sigma_{\rm FUV} & = & N_{\rm FUV} \mu  \\
 & \sim & 0.07 \left( \frac{L_{\rm
       FUV}}{10^{30} \, {\rm erg}\, {\rm s}^{-1}} \right)^{1/2} \nonumber \\
& & \times \left( \frac{3 \,{\rm AU}}{a} \right)^{1/2} \left( \frac{10^{-4}}{f_{\rm T}} \right) \, {\rm g} \, {\rm cm}^{-2}\,. \nonumber
\end{eqnarray}
In Equation (\ref{eq_fuv}), we have normalized $f_{\rm T}$ to its highest plausible value,
appropriate for C. An MRI-active surface density $\Sigma_{\rm{FUV}} \sim 0.07$ g cm$^{-2}$ is
modest, and would drive mass accretion rates only barely observable.
Lowering $f_{\rm{T}}$ would increase $\Sigma_{\rm{FUV}}$. But accounting for extinction
of FUV radiation by dust and molecular hydrogen would decrease
$\Sigma_{\rm{FUV}}$. We are currently undertaking a more careful study of FUV
ionization to quantify these effects.

\acknowledgements Neal Turner provided invaluable feedback during
formative stages of this work. We thank Xue-Ning Bai, Kees Dullemond,
Josh Eisner, Vincent Geers, Al Glassgold, Uma Gorti, Lee Hartmann,
David Hollenbach, Anders Johansen, Yoram Lithwick, Dimitri Semenov,
Greg Sloan, Jim Stone, Marten van Kerkwijk, Yanqin Wu, and Andrew
Youdin for discussions.  Xue-Ning Bai and Jim Stone provided
encouraging feedback that led to additional analyses such as that in
section \ref{sec_timescale}. Zhaohuan Zhu and Lee Hartmann generously
shared their preprint which impressed upon us the need for grain
growth to explain the low optical depths of transitional disk
holes.  An anonymous referee provided a thoughtful and thorough report
that alerted us to the possibility of ``sideways cosmic-rays,'' and
that motivated us to consider the effects of UV ionization.
We also thank our editor, Eric Feigelson, for additional comments on
our manuscript.
E.C. acknowledges the hospitality of the Kavli Institute for Astronomy
and Astrophysics in Beijing, China, where a portion of this work was
carried out.  This work was funded by the National Science
Foundation, in part through a Graduate Research Fellowship awarded to
D.P.-B.

\vspace{0.3in}

\textit{Note added in proof.}--- In a private communication, Xue-Ning Bai
reports that in new, unstratified MRI simulations, the Shakura-Sunyaev
transport parameter $\alpha$ is at most $\sim$$10^{-3}$ when $Am \sim
1$. Combining his result with the results of our paper, we estimate
that the mass accretion rate in the surface layer of a conventional
disk at 3 AU is $\sim$$10^{-11} M_{\odot}$ yr$^{-1}$. At the rim of a
transitional disk, we would predict an accretion rate that is lower by
a factor of $\sim$$h/a$, or $\sim$$10^{-12} M_{\odot}$ yr $^{-1}$ at 3 AU.
These theoretical accretion rates are too low
to explain the observed accretion rates of most disks. The situation
is similar at 30 AU, where the surface layer accretion rate in
conventional (transitional) disks can only be as high as
$\sim$$10^{-9} (10^{-10}) M_{\odot}$
yr$^{-1}$ if Galactic cosmic-rays can penetrate the disk edge-on (see
our Figure 10, right).

\end{document}